\documentclass[10pt, one column, draft cls]{ieeeconf}
\IEEEoverridecommandlockouts 

\usepackage{amsmath,amssymb,bbm,color,psfrag,amsfonts,mathtools}
\usepackage{dsfont}
\usepackage{graphicx,cuted}                       
\usepackage[pdftex,pdfpagelabels,bookmarks,hyperindex,hyperfigures]{hyperref}
\usepackage{wrapfig}
\usepackage{pdfpages}
\usepackage[ruled,lined]{algorithm2e} 
\usepackage{algpseudocode}
   \SetKwInOut{Input}{input}\SetKwInOut{Output}{output}
\usepackage{accents}
\usepackage{lipsum}
\usepackage[T1]{fontenc}
\usepackage[utf8]{inputenc}
\usepackage{mathtools}
\usepackage{multicol}
\usepackage{paralist}   
\usepackage{comment}
\usepackage{kbordermatrix}
\usepackage{tikz}

\usepackage{booktabs}  
\usepackage{multirow}
\usepackage[normalem]{ulem}

\usepackage{caption}
\usepackage{subcaption}

\usepackage{cite}

\newcounter{problem}

\newtheorem{remark}{Remark}

\usepackage{color}

\newcommand{\real}{\mathbb{R}}

\newcommand{\mc}{\mathcal}

\newcommand{\ksmarginphantom}[1]{}
\newcommand{\yzmarginphantom}[1]{}

\newcommand{\setdef}[2]{\left\{#1 \; | \; #2\right\}}

\newcommand{\prb}{\mu_0}

\newcommand{\congfunc}{\ell}

\newcommand{\prior}{\mu_0}
\newcommand{\state}{\omega}
\newcommand{\allstates}{\Omega}
\newcommand{\interior}{\mathrm{int}}
\newcommand{\signal}{\pi}

\newcommand{\pfrac}{\nu}
\newcommand{\simplex}{\mc P}

\newcommand{\nlinks}{n}
\newcommand{\npaths}{\nlinks}

\newcommand{\numsubjects}{S}
\newcommand{\numscenarios}{K}
\newcommand{\Rmax}{r_{\max}}

\graphicspath{{../},{../results/} , {../../../../fig/},{../fig/}}

\algnewcommand{\algorithmicgoto}{\textbf{go to}}%
\algnewcommand{\Goto}[1]{\algorithmicgoto~\ref{#1}}%

\usepackage{tikz}
\usepackage{standalone}
\usepackage{pgfplots}
\usetikzlibrary{positioning}
\usetikzlibrary{shapes, arrows.meta, arrows}
\usepackage{relsize}
\usetikzlibrary{positioning}
\usetikzlibrary{arrows}
\usepackage{verbatim}
\usepackage{soul}

\title{An Experimental Study on Learning Correlated Equilibrium \\ in Routing Games}

\author{Yixian Zhu\thanks{The authors are with the University of Southern California, Los Angeles, CA. \texttt{\{yixian,ksavla\}@usc.edu}. 
K. Savla has financial interest in Xtelligent, Inc.} \qquad Ketan Savla}

\date{\today}

\begin{document}
\maketitle

\begin{abstract}
We study route choice in a repeated routing game where an uncertain state of nature determines link latency functions, and the agents receive private route recommendation. The state is sampled in an i.i.d. manner in every round from a publicly known distribution, and the recommendations are generated by a randomization policy whose mapping from the state is also known publicly. In a one-shot setting, the agents are said to \emph{obey} recommendation if it gives the smallest travel time in a posteriori expectation. A plausible extension to repeated setting is that the likelihood of following recommendation in a round is related to \emph{regret} from previous rounds. If the regret is of satisficing type with respect to a default choice and is averaged over past rounds and over all agents, then the asymptotic outcome under an obedient recommendation policy is known to coincide with the one-shot outcome. We report findings from a related experiment with one participant at a time engaged in repeated route choice decision on computer. In every round, the participant is shown travel time distribution for each route, a route recommendation generated by an obedient policy, and a numeric rating suggestive of average experience of previous participants with the quality of recommendation. Upon entering route choice, the actual travel times derived from route choice of previous participants are revealed. The participant uses this information to evaluate the quality of recommendation received for that round and enter a numeric review accordingly. This is combined with historical reviews to update rating for the next round. Data analysis from 33 participants each with 100 rounds suggests moderate negative correlation between the display rating and the average regret with respect to optimal choice, and a strong positive correlation between the rating and the likelihood of following recommendation. Overall, under obedient recommendation policy, the rating seems to converge close to its maximum value by the end of the experiments in conjunction with very high frequency of following recommendations. 

\end{abstract}

\section{Introduction}
Route choice decision making in traffic networks with travel time uncertainties is a challenging task for drivers. The uncertainties, e.g., due to traffic incidents, are hard to predict and even if known are hard to incorporate optimally into real-time decision-making by the drivers. Therefore, route recommendation platforms are compelling options in such settings. The information advantage of such platforms, i.e., almost real time information about the incidents, combined with availability of huge computational resources, and the ability to make private recommendations, e.g., through mobile devices, also gives these platforms the possibility to steer the driver decisions towards a desired objective. The efficacy however depends on the extent to which the drivers follow the recommendations. This aspect is ignored in typical route recommender systems, e.g., see \cite{Wang.Li.ea:14,Herzog.Massoud.ea:17}.
\emph{Information design}, e.g., see \cite{Bergemann.Morris:19}, provides a framework to formalize the act of following recommendation through the notion of \emph{obedience}. A \emph{private signal}, i.e., mapping from the state to distribution over private recommendations is called obedient in a one-shot game, if following the recommendation leads to smallest travel time in expectation a posterior. The corresponding outcome in the form of link flows is the correlated equilibrium induced by the signal.  

A plausible extension of obedience to repeated setting is that the likelihood of following recommendation in a round is related to \emph{regret} from previous rounds. If the regret is of satisficing type with respect to a default choice and is averaged over past rounds and over all agents, then the asymptotic outcome under an obedient recommendation policy is known to coincide with the one-shot outcome~\cite{Zhu.Savla:DGAA21}. Moreover, this \emph{learning rule} does not require explicit knowledge of the signaling policy by the agents. However, there has been no experimental study of such learning rules or even of correlated equilibrium for non-atomic games to the best of our knowledge. 

There has been a lot of theoretical work on learning Nash equilibria in the context of selfish routing, e.g., see \cite{Fischer.Racke.ea:10,Blum.Even-Dar.ea:10,Krichene.Drighes.ea:15}. There has also been experimental investigation on convergence of route choice behavior to Nash equilibrium, e.g., see \cite{Iida.Akiyama.ea:92,Selten.Chmura.ea:07}. However, corresponding work on correlated equilibria are lacking. In face, existing experimental studies on correlated equilibrium are almost exclusively for 2 X 2 settings and concern primarily Game of Chicken and the Battle of Sexes, e.g., see \cite{Cason.Sharma:07,Duffy2010,Bone.Drouvelis.ea:12,Duffy2017,Arifovic.Boitnott.ea:19}. These studies have identified the following to be key drivers for ensuring high rate of following recommendation: (i) clear and common understanding of the signal which generates the recommendations, often achieved by public announcement of the signaling policy; (ii) the induced correlated equilibrium being Pareto improvement over Nash equilibria; (iii) accurate knowledge of mapping between utility and payoffs of players; (iv) trust that the opponent will follow the recommendation; and (v) fixed vs. random opponent in different rounds, with the former allowing development of strategy profile of the opponent and using it for choice of action. Other studies include \cite{Anbarci2018} which examined the effect of payoff asymmetry on a participant's willingness to follow recommendations.

Some of the considerations mentioned above in the context of following recommendations become more significant for non-atomic games. 
Signaling policies for these games comprise probability distributions with continuous support. This complexity makes the announcement of signal by the mediator as well as its comprehension by the participants to be challenging. This may potentially aggravate known issue with participants not able to execute Bayes rule accurately even for probability mass functions. The issue of trust that other participants will follow recommendations takes assumes a bigger role in the presence of continuum of participants. Furthermore, practical limitations on laboratory experiments necessitates consideration of a pseudo-non-atomic setup, consisting of mixture of a small number of human subjects and a large number of simulated agents. We performed experiments in the limiting case of one human subject at a time with change of participant after a fixed number of rounds.


The outline of our experiments is as follows. One participant at a time engages in repeated route choice decision on computer. In every round, one of five states is i.i.d. sampled according to a probability distribution. The participant is shown travel time forecasts for each of the three routes for the five possible underlying states, a route recommendation generated by an obedient policy conditional on the state realized, and a numeric rating suggestive of average experience of previous participants with the quality of recommendation they received during rounds with the same realized state. The travel time forecasts are based on the assumption that the fraction of other agents who follow recommendation is proportional to the displayed rating, and those who do not follow mimic empirical route choice collected from previous participants. Once the participant enters route choice, the actual travel times on the routes are revealed. These are the  forecasted travel times for the specific state value associated with the round. 
The participant uses this information to evaluate the quality of recommendation received in the round and enters a numeric review accordingly. This is combined with historical reviews to update rating for the next round. 
%
At the conclusion of all its rounds, the participant is asked to fill out a feedback survey, which is designed to get insight into the participant's decision making strategy. Our strategy to announce payoffs, i.e., travel times, is equivalent to announcing the signal. Announcing \emph{forecasted} payoffs based on actual decisions by previous participants, instead of \emph{equilibrium} payoffs in conjunction with rating is meant to elicit trust, and hence recommendation following from the participants. 
Data from 33 participants each with 100 rounds was analyzed to investigate the correlation between empirical probability of following recommendation by the participants, regret, as well as the displayed rating, and the convergence of displayed rating and the empirical distribution of route choice by the participants. 


Our main contributions and findings are as follows. First, we provide an experiment protocol template for studying correlated equilibrium in non-atomic games. Second, through our experimental findings, we provide an instantiation of a setting in which the long run route choice decisions are highly consistent with private recommendations generated by an obedient policy. 
Our third contribution is in verifying elements of a prior dynamic model for obedience in multi-round setting. Our experiments found strong correlation between following recommendation in a particular round and the display rating in that round. We found moderate correlation overall between the review submitted by a participant in a round and the regret which could be associated with the outcome of the round. This correlation however became strong when restricted to participants who demonstrated and self-reported to carefully assess all the information for their route choice decision and for providing review feedback. 
This is important because the notion of regret has been a mainstay in prior theoretical studies on convergence to correlated equilibrium. Overall, the displayed rating converged to maximum value in conjunction with very high likelihood of following recommendation. 

The rest of the paper is organized as follows. Section~\ref{sec:theory} gives the necessary background on modeling and analysis of non-atomic routing game which forms the basis for experiment design. Section~\ref{sec:procedure} provides a general overview of the experiment protocol, while connection with the non-atomic setup and hypotheses are presented in Section~\ref{sec:adapting}. Experimental findings are presented in Section~\ref{sec:findings} with an emphasis on answering the hypotheses. Finally, we present discussion of our results and provide directions for future work in Section~\ref{sec:conclusion}. 

We end this section by defining key notations to be used throughout this paper. Let $\triangle(X)$ denote the set of all probability distributions on $X$. 
For an integer \( n \), we let \( [n]:= \{1, 2, \ldots, n\} \). For $\lambda \geq 0$, let $\simplex_n:=\setdef{x \in \real_{\geq 0}^n}{\sum_{i \in (n)} x_i = 1}$ be the $(n-1)$-dimensional probability simplex.

\section{Theoretical Background}
\label{sec:theory}
We start by providing summarizing model and results from our previous work in \cite{Zhu.Savla:DGAA21} which forms the basis of hypotheses to be tested in the experiments. 

Consider a network consisting of $\nlinks$ parallel links between a single source-destination pair. Without loss of generality, let the agent population generate a unit volume of traffic demand. The link latency functions $\congfunc_{\omega,i}(f_i)$, $i \in [\nlinks]$, give latency on link $i$ as a function of flow $f_i$ through them, conditional on the \emph{state} of the network $\state \in \allstates=\{\state_1, \ldots,\state_{|\Omega|}\}$. The latency function is in the following form:
\begin{equation}
\label{eq:polynomial-latency-function}
\congfunc_{\omega,i}(f_i) =  \sum^D_{d=0} \alpha_{d,\omega,i} \, f_i^d, \, i \in [\nlinks], \, \, \state \in \allstates
\end{equation}
Let $\state \sim \prior \in \interior(\triangle(\allstates))$, for some prior $\prior$ which is known to all the agents. The agents do not have access to the realization of $\state$, but a fixed fraction $\pfrac \in [0,1]$ of the agents receives private route recommendations conditional on the realized state (in the experiments, we have $\nu = 1$, i.e. all agents reveive recommendations). These conditional recommendations are generated by a \emph{diagonal atomic signal} $\signal=\{\signal_{\state} \in \simplex_{\nlinks}: \, \state \in \allstates\}$~\cite{Zhu.Savla:TCNS22}.

Consider the following repeated game setting. The strategy space of  every agent is $\{\mathrm{follow}, \mathrm{do \, not \, follow}\}$. The $\mathrm{do \, not \, follow}$ behavior of the agents is modeled by the row-stochastic matrix $P$ with zeros on the diagonal, where $P_{ij}$ is the fraction of $\mathrm{do \, not \, follow }$ agents who are recommended $i$ but choose $j$. The choice of a agent is driven by a notion of regret. At the end of stage $k$, every agent computes difference between the payoffs associated with the default choice and the recommended choice. These payoff differences are then aggregated over the entire agent population to give $u(k)$. The average of the initial condition $m(1)$ and $u(1), \ldots, u(k)$, denoted as $m(k+1)$, is then mapped to regret as $\theta(m(k+1)) \in [0,1]$, which equals the fraction of agents who do not follow recommendation in stage $k+1$. The details for each step in this process is provided next. 

We assume that upon completion of the trips, all the agents have access to traffic report from the $k$-th stage. This report consists of $\omega(k)$ and $\{\ell_{\omega(k),i}\}_{i \in [n]}$. For simplicity, first consider the two-link case with $\ell_{\omega(k),1}>\ell_{\omega(k),2}$. For an agent who was recommended route 1, irrespective of whether she follows the recommendation or not, her recommendation is sub-optimal by $\ell_{\omega(k),1}-\ell_{\omega(k),2}$. In the general case of $n \geq 2$ links, for an agent who follows recommendation to take route $i$, the sub-optimality of the recommendation is $\congfunc_{\omega(k),i} -\sum_{j \in [n]} P_{ij} \congfunc_{\omega(k),j}$. On the other hand, for an agent who does not follow the recommendation to take route $i$ but rather takes route $j$, the sub-optimality of her recommendation is $\congfunc_{\omega(k),i} -\congfunc_{\omega(k),j}$. Taking into account the number of agents who are recommended different routes and the fraction of them who $\mathrm{follow}$ or $\mathrm{do \, not \, follow}$ the recommendation, the aggregation of payoff difference over the entire agent population is given by:
\begin{align}
u(k) := & \sum_{i \in [n]} \Big(\congfunc_{\omega,i} -\sum_{j \in [n]} P_{ij} \congfunc_{\omega,j} \Big) \pi_{\omega,i} \left(1-\theta(m(k))\right) + \sum_{i,j \in [n]} \Big(\congfunc_{\omega,i} - \congfunc_{\omega,j} \Big) P_{ij}\pi_{\omega,i} \theta(m(k)) \nonumber \\
=& \sum_{i,j \in [n]} \Big(\congfunc_{\omega,i} - \congfunc_{\omega,j} \Big) P_{ij} \pi_{\omega,i} 
= {\pi_{\omega}}^T (I-P) \congfunc_{\omega}
\label{eq:instant} 
\end{align}
where we have dropped the dependence of $\{\ell_{\omega,i}\}_{i \in [n]}$, $\{\pi_{\omega,i}\}_{i \in [n]}$ and $\omega$ on $k$ for brevity. 
The average of these instantaneous payoff differences (as well as the initial condition $m(1)$) is:
\begin{equation}
\label{eq:S-agent-update}
\begin{split}
m(k+1) & = \frac{1}{k+1} \left(m(1)+ u(1) + \ldots u(k) \right) = \frac{k}{k+1} m(k) + \frac{1}{k+1} u(k)
\end{split}
\end{equation}

We adopt the following notion of regret:
\begin{equation}
\label{eq:rating-to-mistrust}
\theta(m(k))=\frac{[m(k)]^+}{m_{\max}}
\end{equation}
where $m_{\max}$ is chosen to be sufficiently large so that $\theta(m(k)) \in [0,1]$ for all possible values of $m(k)$. Since $[m(k)]^+$ can be upper bounded by $\sum_{i \in [n]} \sum_{ d=0}^D \max_{\omega \in \Omega} \alpha_{d,\omega,i}$, it can then serve as a lower bound for $m_{\max}$.

$\theta(m(k))$ is interpreted as the fraction of agents who do not follow recommendation. 
Therefore, the link flows induced by the agents is given by: \ksmarginphantom{need more details for \eqref{eq:p flow dynamics-m-general}}
\begin{equation*}
\label{eq:p flow dynamics-m-general}
\begin{split}
f_i(m(k),\omega(k)) = \pi_{\omega(k),i} \left(1-\theta(m(k))\right) + \sum_{j \in [n]} P_{ji} \pi_{\omega(k),j} \theta(m(k))
\end{split}
\end{equation*}
which in matrix form becomes:
\begin{align}
f(m(k),\omega(k)) & = \pi_{\omega(k)} \left(1-\theta(m(k))\right) + \sum_{j \in [n]} P_{ji} \pi_{\omega(k),j} \theta(m(k)) = \pi_{\omega(k)} - \theta(m(k)) \pi_{\omega(k)} + \theta(m(k)) P^T \pi_{\omega(k)}  \nonumber \\
& = \pi_{\omega(k)} + \theta(m(k)) (P^T - I ) \pi_{\omega(k)}
\label{eq:p flow dynamics-m-general-matrix}
\end{align}

\begin{remark}
The framework in \eqref{eq:instant}-\eqref{eq:p flow dynamics-m-general-matrix} is reminiscent of the \emph{regret matching} framework of \cite{Hart.Mas-Colell:00}. A few points are worth special emphasis:
\begin{enumerate}
\item The aggregation of payoff difference over all agents in \eqref{eq:instant} is inspired by platforms such as Yelp. 
\item In the framework of \cite{Hart.Mas-Colell:00}, $[m]^+$ is referred to as regret and $m_{\max}$ is the inertia parameter which captures the propensity of an agent to stick to the action choice in the previous round. On the other hand, in our setup, $\theta=[m]^+/m_{\max}$ is interpreted as the degree of obedience of an individual agent, tuned by the parameter $m_{\max}$.
\end{enumerate}
\end{remark}

We consider $\signal$ which are \emph{obedient}, i.e., which satisfy
$\sum_{\state} \ell_{\omega,i}(\signal_{\state,i}) \, \signal_{\state,i} \, \prior(\state) \leq \sum_{\state}  \ell_{\omega,j}(\signal_{\state,j}) \, \signal_{\state,i} \, \prior(\state)$ for all $i, j \in [\npaths]$. We showed in \cite{Zhu.Savla:DGAA21} that, for any $\prb$, obedient $\pi$, and every $m(1) \in [-m_{\max},+m_{\max}]$, the solution to \eqref{eq:instant}-\eqref{eq:p flow dynamics-m-general-matrix} satisfies $\lim_{k \rightarrow \infty} f_i(k) - \pi_{\omega(k),i} = 0$ for all $i \in [n]$.

\section{Experiment Procedure}
\label{sec:procedure}
Simulating non-atomic setup requires simultaneous participation by a very large number of participants. Practical limitations on laboratory experiments therefore necessitate consideration of a pseudo-non-atomic setup, consisting of mixture of a small number of human participants and a large number of simulated participants. We performed experiments in the limiting case of one human participant at a time with change of participant after a fixed number of rounds.

The experiment protocol was reviewed and approved by the Institutional Review Board at the University of Southern California (USC \# UP-22-00107). 
The experiment was conducted in the Kaprielian Hall at USC during April 2022 and May 2022. A total of 34 participants with a good command of the English language and with no prior experience with our experiment were recruited from the undergraduate population of the university. Upon arrival at the laboratory, participants were given a presentation by an experiment personnel, aided by slide illustrations, available at \cite{RoutingExpSlides}. 
The experiments was run on a networked desktop computer, and programmed using Python. \yzmarginphantom{Asked Christine for detailed info regarding the programming software and database, waiting for the reply.} Except for experiment personnel, only one participant was present in the laboratory during an experiment session. 

\begin{figure}[htb!]
\begin{center}
\begin{minipage}[c]{.475\textwidth}
\begin{center}
\includegraphics[width=1.0\textwidth]{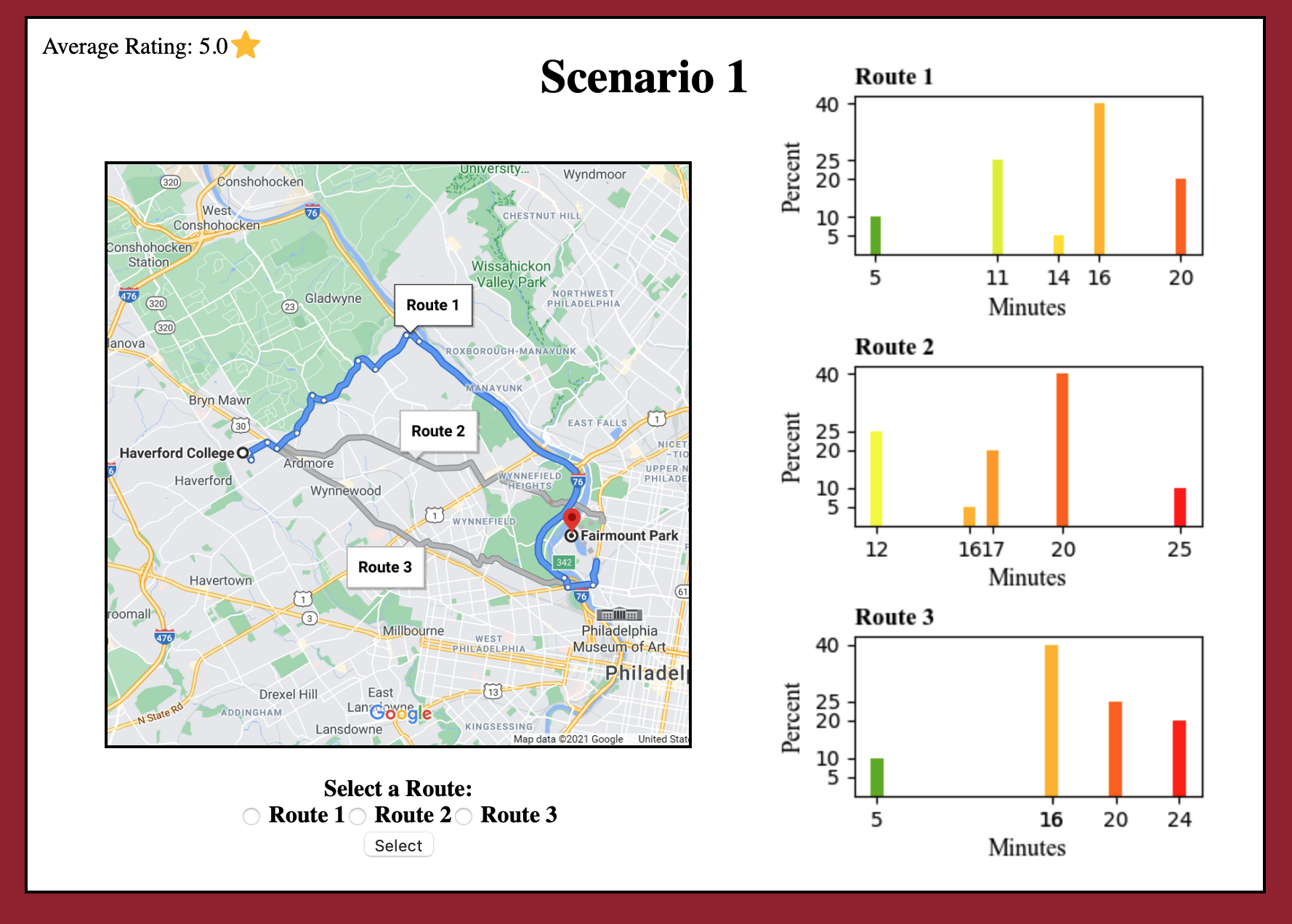} 
\end{center}
\end{minipage}
\end{center}
\caption{\sf User interface before route selection during a typical scenario.}
\label{fig:user-interface-1}
\end{figure}

\begin{figure}[htb!]
\begin{center}
\begin{minipage}[c]{.475\textwidth}
\begin{center}
\includegraphics[width=1.0\textwidth]{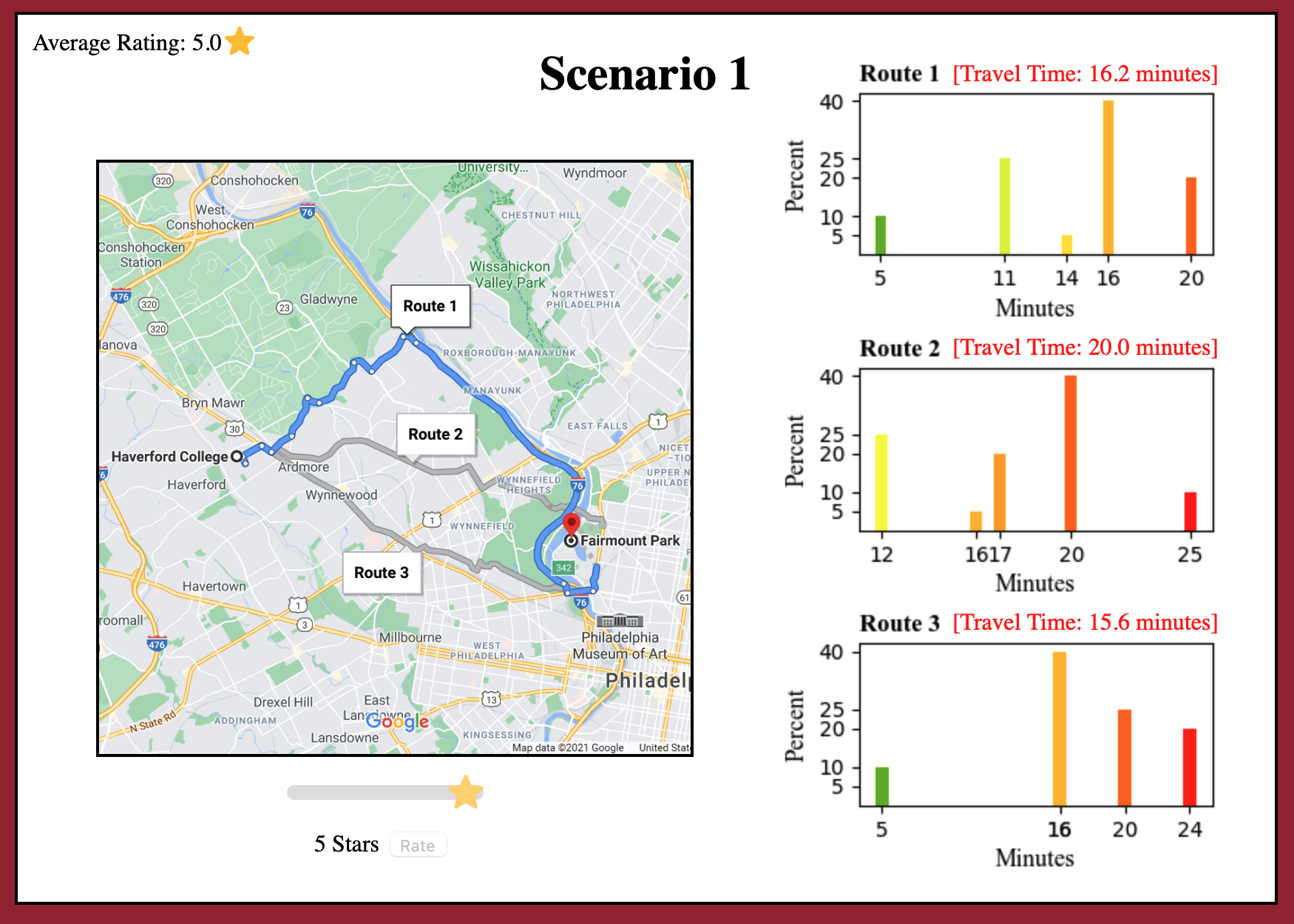} 
\end{center}
\end{minipage}
\end{center}
\caption{\sf User interface after route selection and before review submission during a typical scenario.}
\label{fig:user-interface-2}
\end{figure}

During the experiment, a participant had to select one among three routes in a traffic network on computer, in a series of rounds or \emph{scenarios}. The traffic network was chosen to be outside the Los Angeles area to minimize bias. 
The information displayed during a typical scenario, as shown in Figure~\ref{fig:user-interface-1}, consisted of: (i) Traffic network with three possible routes to go from the origin to the destination, marked as Route 1, Route 2, and Route 3. The route highlighted is the recommended route; (ii) Travel time forecasts on different routes corresponding to different underlying states, shown via histogram in the right side of the screen; (iii) Average rating at the top left indicating what other participants in the past thought of the quality of recommendation in terms of recommending the fastest path. The average rating is on a scale of 1 to 5 with 5 being the highest; and (iv) Menu at the bottom to enter route selection. During the pre-experiment presentation, the participants were instructed to make route choice based on the information contained in (i)-(iii). Specifically, they were instructed to let their likelihood of following the recommendation be proportional to the displayed rating, and to use the histogram when they decide not to follow the recommendation. Upon selecting a route, and pressing the "Select" button, the computer revealed the \emph{actual} travel time on top of the histogram for each route as in Figure~\ref{fig:user-interface-2}. These are the forecasts for the specific state value associated with the round. 


\ksmarginphantom{check Figures~\ref{fig:user-interface-1} and \ref{fig:user-interface-2} for why the histograms are incomplete}

The travel time forecasts are computed from \emph{forecasted} link flows using the link latency functions in \eqref{eq:polynomial-latency-function}, where the state $\omega$ is predetermined and is the same for all the participants in the same round. The forecasted link flows are computed by assuming that the other agents follow recommendation with a likelihood proportional to the displayed rating, and for those who do not follow, their route choice mimics the alternate choices made by previous participants under same state and same displayed rating; see \eqref{eq:human-output-to-input-1}-\eqref{eq:human-output-to-input-2} for details.
 The participant is required to enter his/her own experience with the quality of recommendation in this scenario by choosing a review rating on a sliding scale at the bottom of the screen. During the pre-experiment presentation, the participants were instructed to consider not only the rankings but also the difference in values of actual travel times of both the recommended route and the route chosen by participants to come up with a rational review rating.
Upon pressing the "Rate" button, the computer showed the next scenario whose displayed rating is the average of the review just submitted by the participant and the reviews submitted by previous participants under same state and same displayed rating; see \eqref{eq:human-output-to-input-1} for details. The process repeats until all the scenarios are done, after which the participant was asked to fill a survey to get insight into his/her route choice decision making strategy during the experiment. This survey is provided in the Appendix.

\ksmarginphantom{need to provide some justification for this choice of performance}
All participants received a show-up renumeration of US \$10 and an additional maximum of US \$10 depending on the \emph{level of homogeneity} that they exhibit. For a given participant, this is measured in terms of the difference between his/her empirical conditional route choice probability distributions and that of the previous participants. This is then mapped proportionally to the renumeration amount. The conditional probability distribution here refers to the probability of choosing route $j$ when recommended route $i$. The additional renumeration averaged \$9.5 across all participants. We used a pretty generation mapping from a participant's level of homogeneity to his/her renumeration, and therefore the high average is not necessarily an indication of high level of homogeneity among the participants.  All the renumeration was paid in cash. 
Participant sessions lasted for approximately an hour on average. 

For each scenario, we recorded the displayed rating (between 0 and 5), the recommended route ($\{1,2,3\}$), the selected route ($\{1,2,3\}$), the submitted review (between 0 and 5), the start and end times (the local clock time of the computer running the experiment program).

\section{Adapting Theory to Experiment Setup and Hypotheses}
\label{sec:adapting}
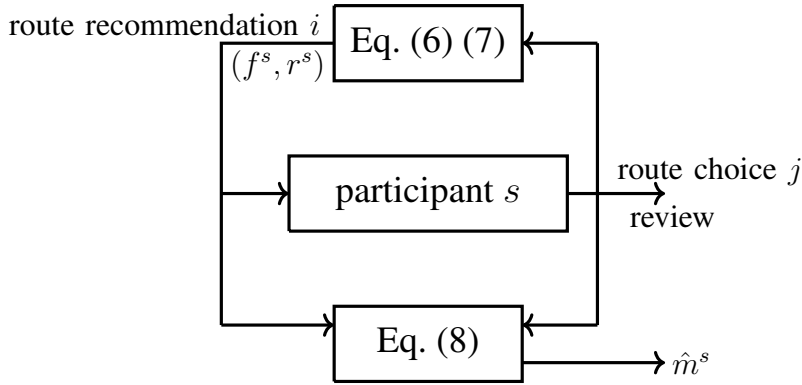
\begin{figure}[htb!]
\begin{center}
\begin{tikzpicture}
\draw[very thick] (-1.85,-0.5) -- (-1.85,0.5) -- (1.85,0.5) -- (1.85,-0.5) -- (-1.85,-0.5);
\draw (0,0) node {\Large participant $s$};

\draw[very thick] (-1.25,-2.5) -- (-1.25,-1.5) -- (1.25,-1.5) -- (1.25,-2.5) -- (-1.25,-2.5);
\draw (0,-2) node {\Large Eq. \eqref{eq:subject-model}};

\draw[very thick] (-1.25,2.5) -- (-1.25,1.5) -- (1.25,1.5) -- (1.25,2.5) -- (-1.25,2.5);
\draw (0,2) node {\Large Eq. \eqref{eq:human-output-to-input-1} \eqref{eq:human-output-to-input-2}};
\draw (3.5,-2.25) node {\large $\hat{m}^s$};

\draw (-3.5,2.25) node {\large route recommendation $i$};
\draw (-2,1.7) node {\large $(f^s,r^s)$};

\draw[very thick,->] (1.85,0) -- (3.15,0);
\draw[very thick,->] (1.25,-2.25) -- (3.15,-2.25);

\draw[very thick] (2.25,0) -- (2.25,-1.75);
\draw[very thick,->] (2.25,-1.75) -- (1.25,-1.75);

\draw[very thick] (2.25,0) -- (2.25,2.0);
\draw[very thick,->] (2.25,2.0) -- (1.25,2.0);

\draw[very thick] (-1.25,2.0) -- (-2.75,2.0) -- (-2.75,-1.75);
\draw[very thick,->] (-2.75,0) -- (-1.85,0);
\draw[very thick,->] (-2.75,-1.75) -- (-1.25,-1.75);

\draw (3.75,0.3) node {\large route choice $j$};
\draw (3.25,-0.3) node {\large review};

\end{tikzpicture}
\caption{\sf Input-output illustration for the participant and the simulated model.}
\label{fig:blockdiagram}
\end{center}
\end{figure}

Let there be $\numsubjects$ human participants in total, each participating for $\numscenarios$ rounds. The numbering of the participants is in the order in which they participate. Figure~\ref{fig:blockdiagram} shows that the response from a participant in every round consists of route chosen in that round as well as the review about the quality of recommendation in that round. In Section~\ref{sec:theory}, we provided a model for an estimate of this review to be related to a notion of instantaneous regret. We modify the notion for the experiments as follows. Let $\ell_1, \ldots, \ell_{\nlinks}$ be the realized travel times on the $\nlinks$ links, and let $j$ be the recommended route. Then, we let instantaneous regret be equal to $\ell_j - \min_{i \in [\nlinks]} \ell_i$. 

The responses of the participants are summarily stored and updated as follows:
\begin{itemize}
\item $\mc R(s,k;\omega,r)$: set of the reviews submitted by participants $1, \ldots, s-1$ in all their rounds $1, \ldots, \numscenarios$, and by participant $s$ in rounds $1, \ldots, k$,  when the state realization was $\omega$, and the displayed rating was $r$. Let $\bar{\mc R}(s,k;\omega,r)$ be the average of those review values.
\item $\mc M(s,k;\omega,r)$: set of instantaneous regret for participants $1, \ldots, s-1$ in all their rounds $1, \ldots, \numscenarios$, and by participant $s$ in rounds $1, \ldots, k$, when the state realization was $\omega$, and the displayed rating was $r$. Let $\bar{\mc M}(s,k;\omega,r)$ be the average of these regret values.
\item $\mc N(s;i,j)$: number of times participants $1, \ldots, s-1$ choose route $j$ when recommended route $i$ in all their rounds $1, \ldots, \numscenarios$.
\end{itemize}

Furthermore, the matrix $P$ is intrinsic to the participants and needs to be estimated online during the experiments. We adopt the following natural estimate: 
\begin{equation*}
\hat{P}_{ij}^{s}= \begin{cases}\frac{\mc N(s;i,j)}{\sum_{j \in [\nlinks], j\neq i} \mc N(s;i,j)} & i\neq j \\ 0 & i=j \end{cases} \quad i, j \in [\nlinks], \quad s \in \{2, \ldots, \numsubjects\}
\end{equation*}
starting with a specified $\hat{P}^1$ for the first participant. 

The ratings to be displayed as well as link flows for the human participant $s \in [\numsubjects]$ in various rounds are given by\footnote{The rating $r^s$ is quantized to the nearest first decimal place when displaying to the participants.}: for $k = 1, \ldots, \numscenarios-1$, 
\begin{align}
r^s(k+1) & =  \frac{k}{k+1} r^s(k) + \frac{1}{k+1} \, \bar{\mc R}(s,k;\omega(k),r^s(k)) \label{eq:human-output-to-input-1} \\
f^s_i(k+1) & = \pi_{\omega(k+1),i} \frac{r^{s}(k+1)}{\Rmax} + \sum_{j \in [\nlinks]} \hat{P}^{s}_{ji} \pi_{\omega(k+1),j} \left(1-\frac{r^{s}(k+1)}{\Rmax}\right) \label{eq:human-output-to-input-2}
\end{align}
starting with a specified $r^s(1)$ for the first round. 

We modify the model for human participant from \eqref{eq:instant}-\eqref{eq:S-agent-update} as: $k = 1, \ldots, \numscenarios$,
\begin{equation}
\label{eq:subject-model}
\begin{split}
\hat{u}^s(k) & = \bar{\mc M}(s,k;\omega(k),r^s(k)) \\
\hat{m}^s(k) & = \frac{1}{k} \left(\hat{u}^s(1) + \ldots + \hat{u}^s(k) \right)
\end{split}
\end{equation}

\subsection{Hypotheses}
Our objective is to test the following hypotheses:
\begin{enumerate}
\item[(H1)] Empirical probability of a participant choosing recommended route is proportional to the value of the rating displayed with the recommendation. The findings for this hypothesis are in Section~\ref{sec:hypo1}.
\item[(H2)] The displayed rating, i.e., $r^{s}(\cdot)$, converges to its maximum value $\Rmax$ as $s$ increases. The empirical probability of participants choosing recommended route converges to $1$ as $s$ increases. The findings for this hypothesis are in Sections~\ref{sec:displayed-rating} and \ref{sec:following}.
\item[(H3)] The empirical distributions of route choices when not following recommendation, i.e., off-diagonal entries of $\hat{P}^s$, converge as $s$ increases. The findings for this hypothesis are in Section~\ref{sec:Phat}.
\item[(H4)] There is a negative correlation between time-averaged aggregated regret $\hat{m}^{s}(\cdot)$ and the displayed rating $r^s(\cdot)$. The findings for this hypothesis are in Section~\ref{sec:display-vs-aggregated-regret}.
\end{enumerate}

\section{Experiment Findings}
\label{sec:findings}
\subsection{Experiment Parameters}
The parameters in our experiments were $\nlinks=3$, $|\Omega|=5$,  
$\numsubjects=33$\footnote{34 participants were recruited in total, but one participant faced technical error during experiment, and therefore we discarded the data from this participant from our analysis.}, $\numscenarios=100$, $\hat{P}^1=\begin{bmatrix}0&0&1\\0&0&1\\0.5&0.5&0\end{bmatrix}$, $r^s(1)=2.5$ for all $s \in [\numsubjects]$, $\Rmax=5$, 
%
%
\begin{equation*}
\alpha_0 = 
\kbordermatrix{&i=1&i=2&i=3\\
\omega_1&5&25&4\\
\omega_2&20&15&24\\
\omega_3&15&20&14\\
\omega_4&11&15&16\\
\omega_5&8&10&20}, \quad 
\alpha_1 = 
\kbordermatrix{&i=1&i=2&i=3\\
\omega_1&4&2&1\\
\omega_2&1&2&3\\
\omega_3&2&3&4\\
\omega_4&3&5&2\\
\omega_5&5&4&5}, \quad
\pi = 
\kbordermatrix{&i=1&i=2&i=3\\
\omega_1&0.1&0&0.9\\
\omega_2&0&1&0\\
\omega_3&0.6&0&0.4\\
\omega_4&0.9&0.1&0\\
\omega_5&0.6&0.4&0}
\end{equation*} \ksmarginphantom{need to rather compare link-wise travel times under policy and Wardrop}
The $\pi$ chosen is an optimal solution to the \emph{information design problem}~\cite{Zhu.Savla:TCNS22}, i.e., it minimizes social cost among all obedient policies. Indeed, the social cost under this policy is about 13\% lower than under Bayes Wardrop equilibrium. 
The sequence $\{\omega(k)\}_{k=1}^{100}$ is sampled offline in an i.i.d. manner from $\mu_0=[0.1 \, \, 0.2 \, \, 0.4 \, \, 0.05 \, \, 0.25]$. The empirical distribution of the sampled sequence is equal to the prior $\mu_0$, and the same sampled sequence is used for all the participants. 
 

\subsection{Recommendation Following vs Displayed Rating}
\label{sec:hypo1}
\begin{figure}[htb!]
\begin{center}
\begin{minipage}[c]{.475\textwidth}
\begin{center}
\includegraphics[width=1.0\textwidth]{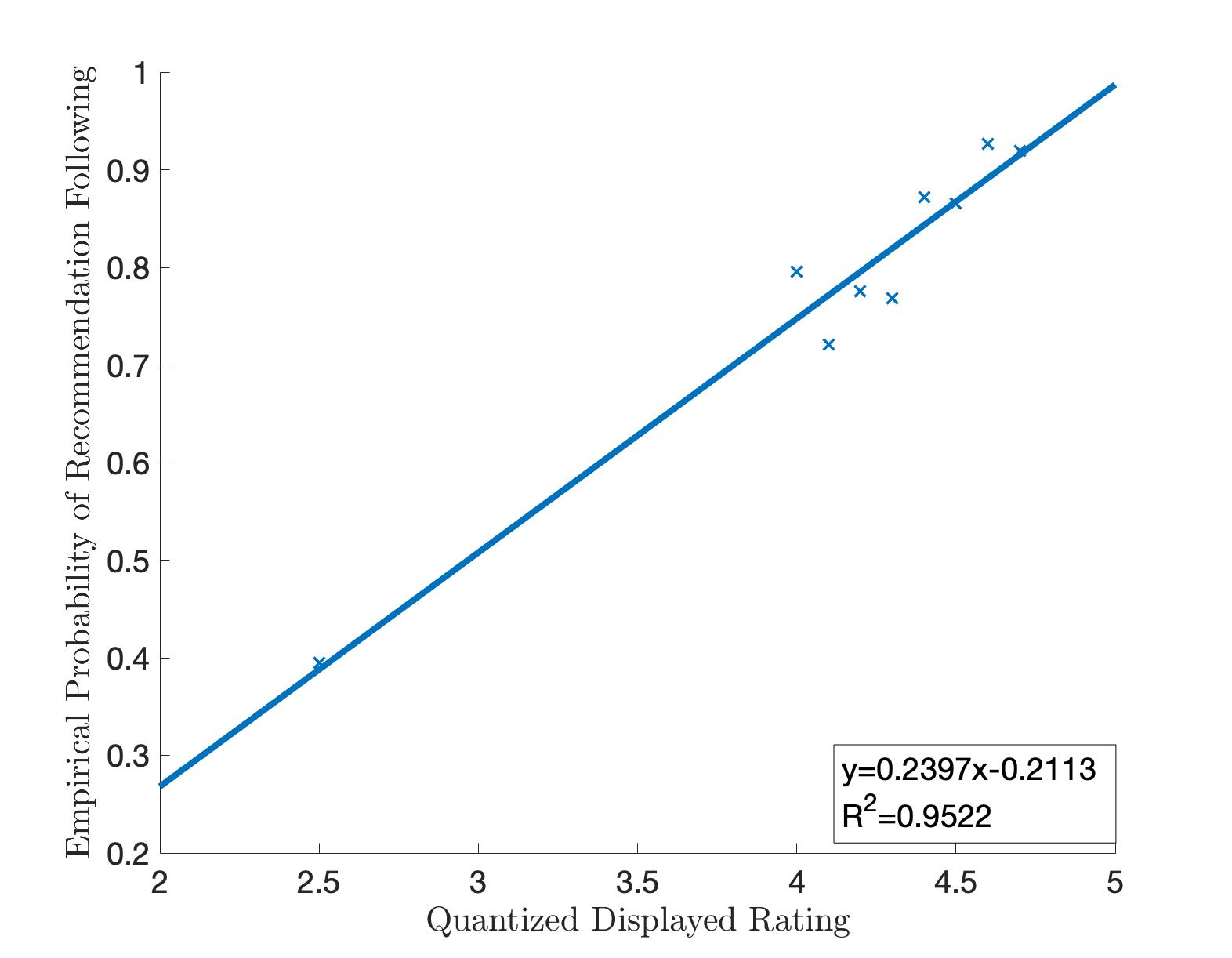} 
\end{center}
\end{minipage}
\end{center}
\caption{\sf Linear regression between empirical probability of recommendation and quantized displayed rating.}
\label{fig:hypo1}
\end{figure}
For the sake of analysis, we quantized the displayed rating by rounding off to one decimal place. We computed empirical probabilities of following recommendation over all participants and all scenarios for each quantized displayed rating. We discarded data associated with the displayed rating interval $[2.6,3.9]$ because the rating changed so rapidly in $[2.6,3.9]$ that the  participants saw a specific display rating in the range $[2.6,3.9]$ for less than one scenario on average. Given such a low sample size, the statistics for following recommendation conditional on display rating in the interval $[2.6,3.9]$ would not be reliable. Moreover, these discarded data points accounted for only 4.8$\%$ of all data points. Figure~\ref{fig:hypo1} shows the outcome of linear regression between the empirical probabilities and displayed ratings. The coefficient of determination $R^2=0.9522$ suggests a strong correlation. The positive slope suggests a positive correlation, and the negative intercept suggests that participants would not follow the recommendations when the displayed rating is less than $\frac{0.2113}{0.2397} \approx 0.9$. In the extreme case when the quantized display rating is $5$, the linear regression model generates a projected empirical probability of $0.9872$ which is consistent with the theoretical model in \eqref{eq:p flow dynamics-m-general-matrix}.

\subsection[]{Long Run Behavior of \(\hat{P}\)}
\label{sec:Phat}
\begin{figure}[htb!]
\begin{center}
\begin{minipage}[c]{.3\textwidth}
\begin{center}
\includegraphics[width=1.0\textwidth]{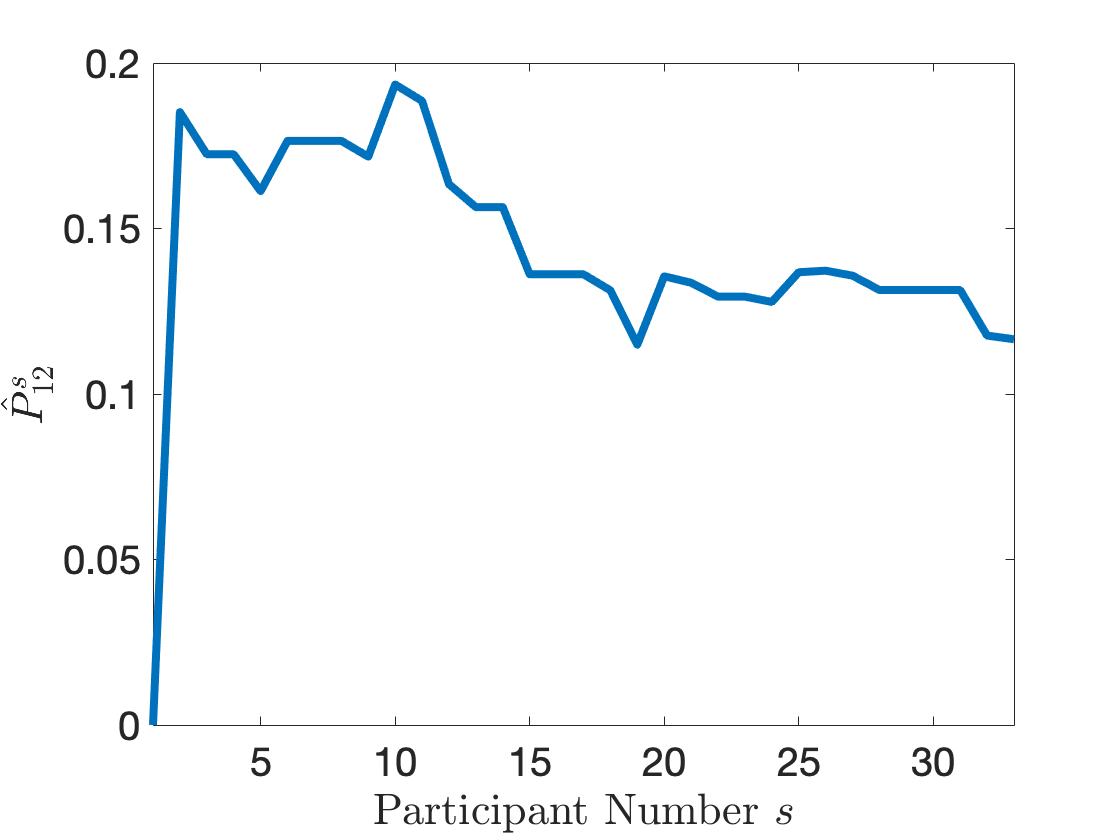} 
\end{center}
\end{minipage}
\begin{minipage}[c]{.3\textwidth}
\begin{center}
\includegraphics[width=1.0\textwidth]{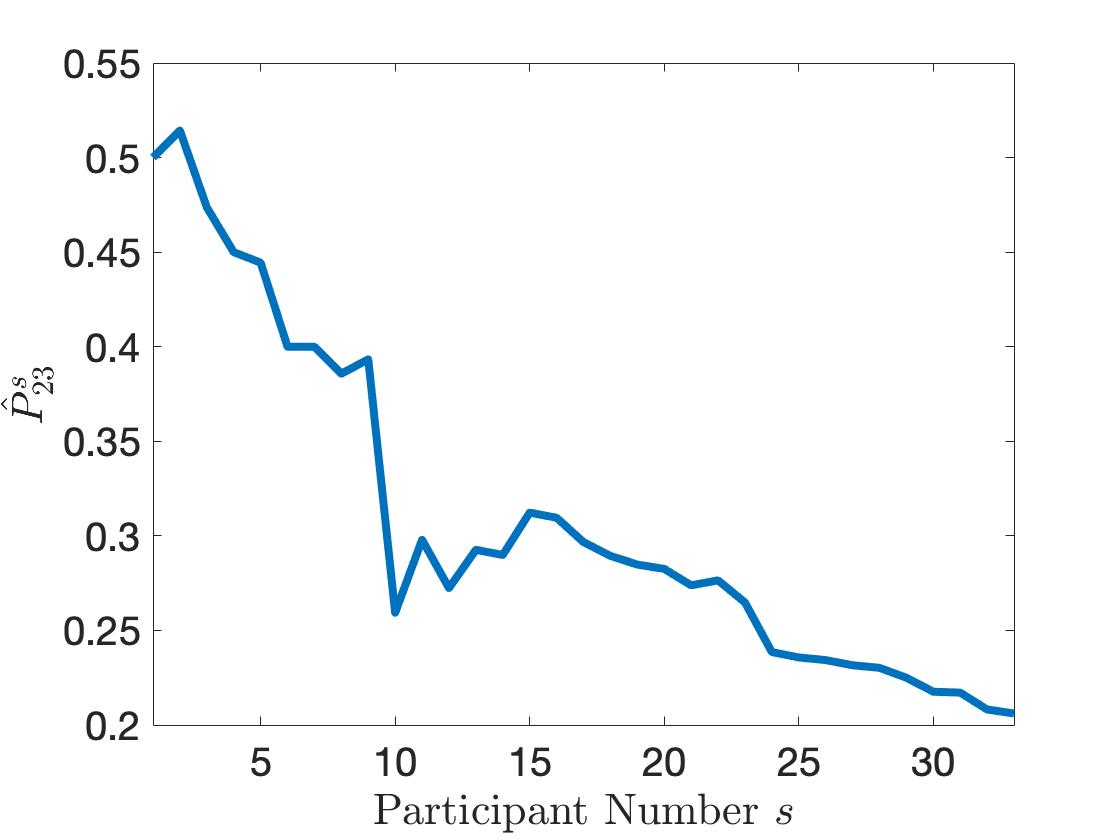} 
\end{center}
\end{minipage}
\begin{minipage}[c]{.3\textwidth}
\begin{center}
\includegraphics[width=1.0\textwidth]{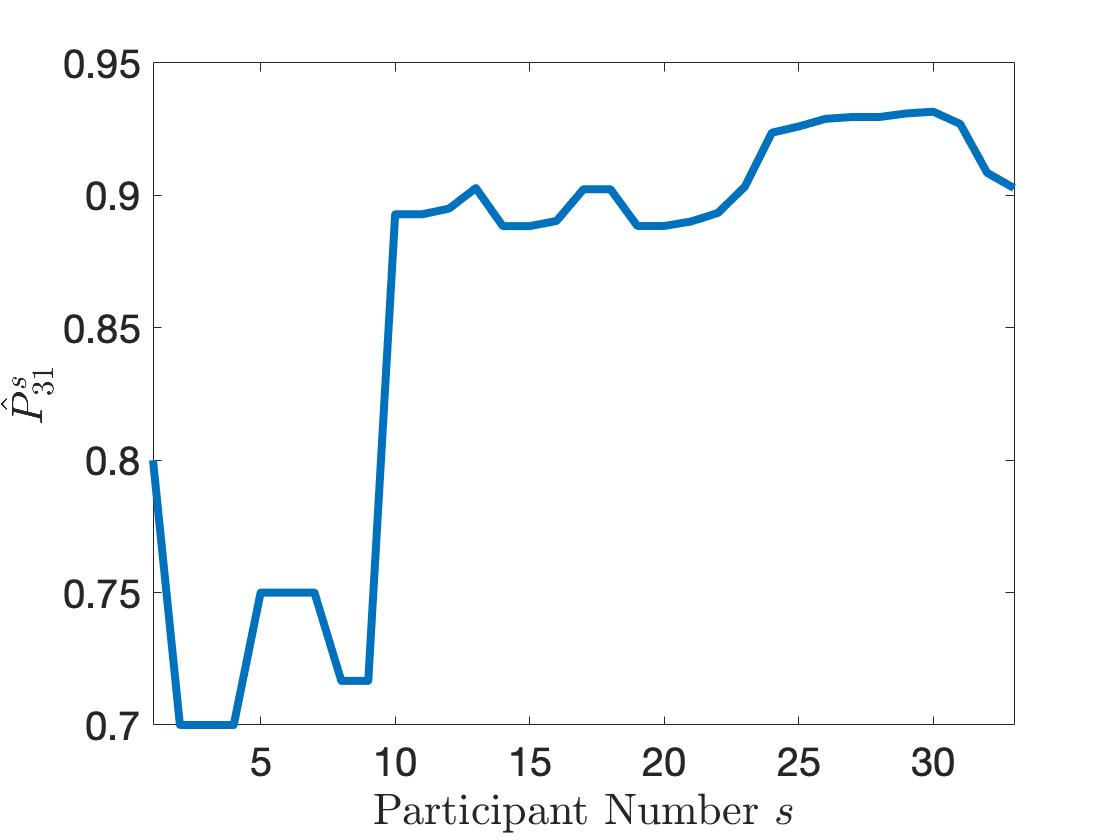} 
\end{center}
\end{minipage}
\end{center}
\caption{\sf Evolution of $\hat{P}^s$ with participant number $s$.}
\label{fig:hypo1_P}
\end{figure}

Recall that $\hat{P}^s$ is held constant throughput the session of participant $s$, and updated to $\hat{P}^{s+1}$ at the end of the session. Also, note from the definition that $\hat{P}^s$ is a row-stochastic matrix, and that its diagonal entries are zero. In our current case of $\nlinks=3$, this therefore leaves $3$ independent entries, say $\hat{P}^{s}_{12}, \hat{P}^{s}_{23}$ and $\hat{P}^{s}_{31}$. The evolution of these quantities with increasing $s$ is shown in Figure~\ref{fig:hypo1_P}. The plots suggest convergence of $\hat{P}^s$ as $s$ increases.

\subsection{Long Run Behavior of the Displayed Rating}
\label{sec:displayed-rating}
\begin{figure}[htb!]
\begin{center}
\begin{minipage}[c]{.475\textwidth}
\begin{center}
\includegraphics[width=1.0\textwidth]{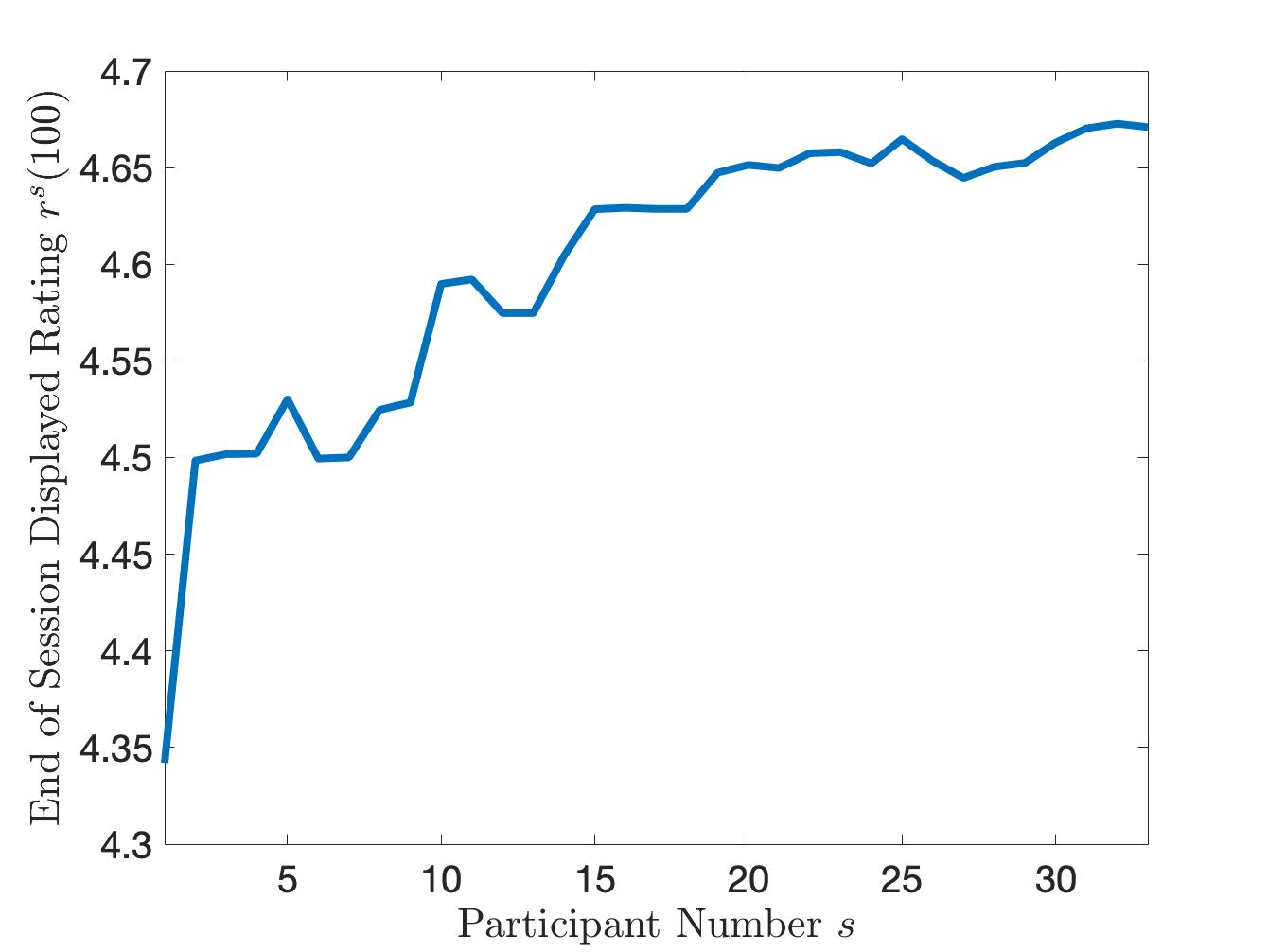} 
\end{center}
\end{minipage}
\end{center}
\caption{\sf Evolution of end of session display rating $r^s(100)$ with participant number $s$.}
\label{fig:hypo2_rating}
\end{figure}

\begin{figure}[htb!]
\begin{center}
\begin{minipage}[c]{.475\textwidth}
\begin{center}
\includegraphics[width=1.0\textwidth]{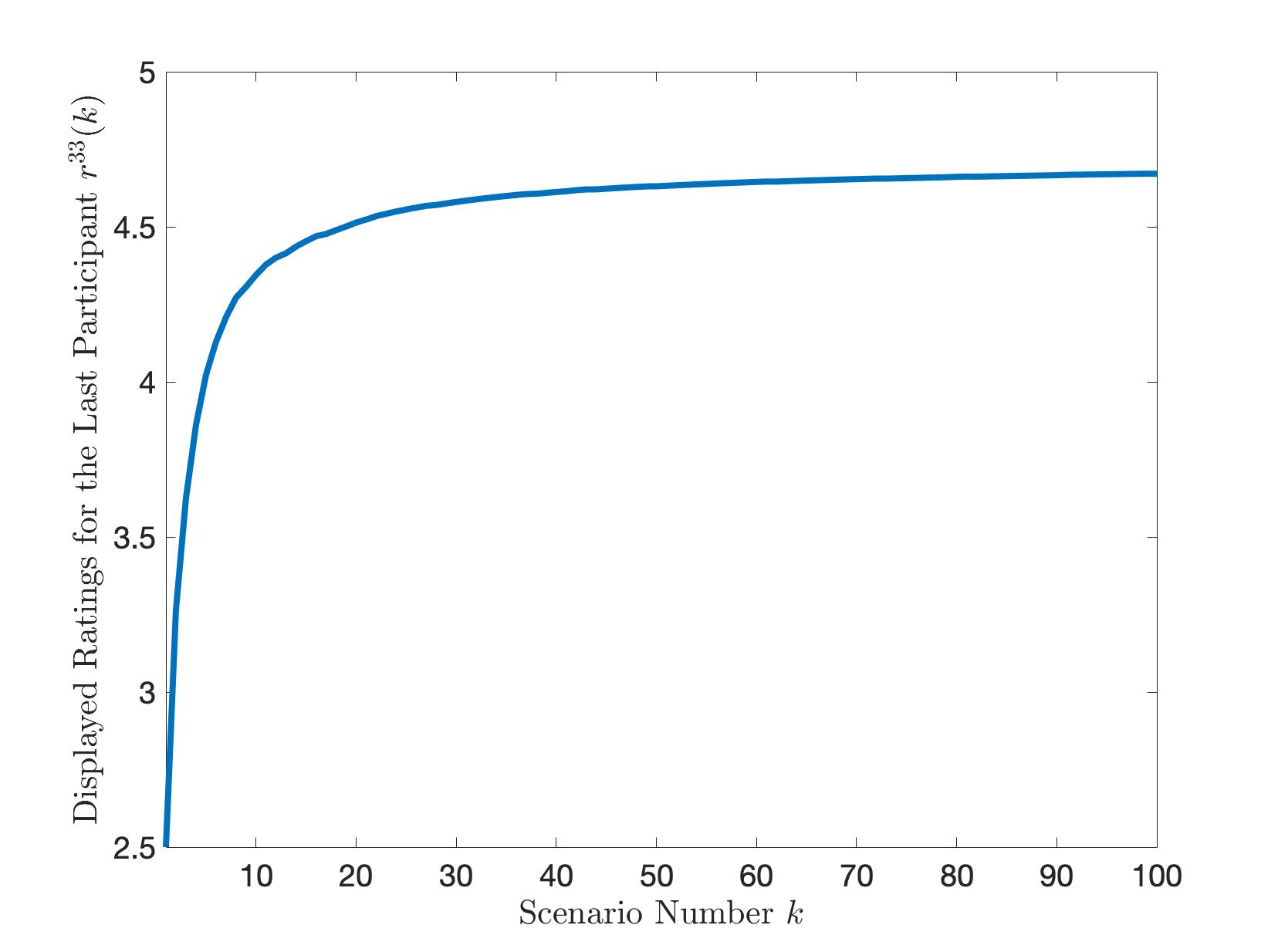} 
\end{center}
\end{minipage}
\end{center}
\caption{\sf Evolution of the display rating during the session of the last participant, i.e., $r^{33}(k)$, with scenario number $k$.}
\label{fig:hypo2_rating_34}
\end{figure}

Recall from \eqref{eq:human-output-to-input-1} that in order to simulate review collection from multiple participants participating simultaneously, we rather collect reviews from previous scenarios by previous participants who made route choice under the same realization of state and for the same displayed rating. Naturally, the accuracy of such a surrogate becomes better with more data collection. Motivated by this, we studied the long run behavior of displayed rating $r^s$ in two ways. First, we studied the end of session displayed rating with increasing participant number. As Figure~\ref{fig:hypo2_rating} illustrates, that this value increases to roughly $4.67$, and seems to be settling around this value, by the last participant. 
Second, we studied the displayed rating during the session of the last participant participant $\#33$. Figure~\ref{fig:hypo2_rating_34} shows a monotonic increase of the displayed rating from $2.5$ (recall that the initial condition is set to be $r^s(1)=2.5$ for all the participants) to a value of around $4.67$ by the end of the session. The evolution of display rating is understandably smoother given that the set $\bar{\mc R}$ used in \eqref{eq:human-output-to-input-1} has sufficient samples for all frequently occurring combinations of $\omega$ and $r$ by the time of the last participant. 


\subsection{Long Run Empirical Probability of Following Recommendation}
\label{sec:following}
\begin{figure}[htb!]
\begin{center}
\begin{minipage}[c]{.475\textwidth}
\begin{center}
\includegraphics[width=1.0\textwidth]{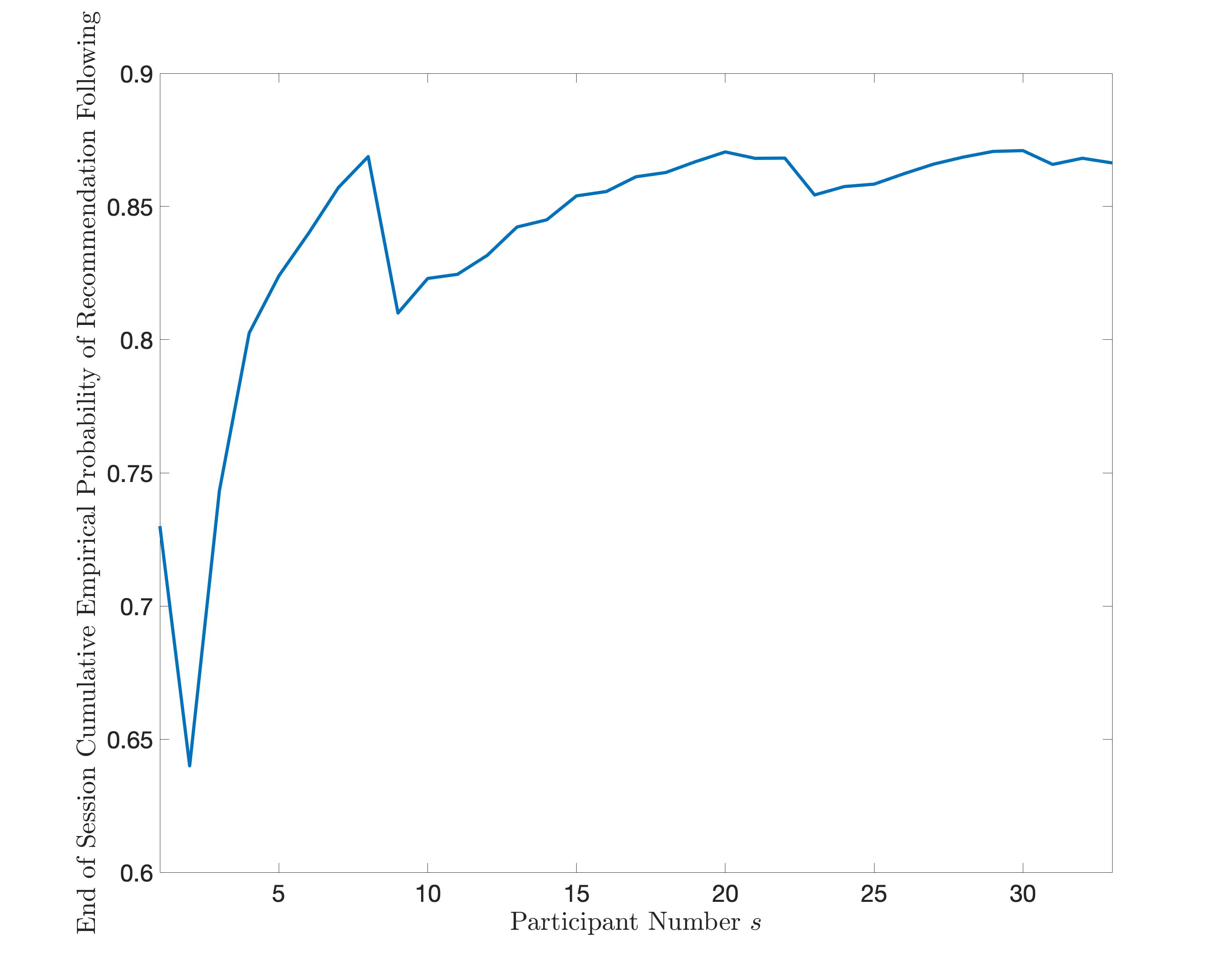} 
\end{center}
\end{minipage}
\end{center}
\caption{\sf Evolution of end of session cumulative empirical probability of following recommendation with increasing participant number.}
\label{fig:hypo2_rate}
\end{figure}

Figure~\ref{fig:hypo2_rate} shows the evolution of cumulative empirical probability of following recommendations from the start of session for participant 1 through the end of session for $s$ for increasing value of $s$. 

This cumulative empirical probability in Figure~\ref{fig:hypo2_rate} seems to converge to around $0.87$. This compares very well with the prediction of the linear regression model of Section~\ref{sec:hypo1}, according to which the empirical probability of following recommendation for the display rating of $4.67$ (the long run value from Section~\ref{sec:displayed-rating}) is approximately equal to $0.2397\times 4.67-0.211=0.91$. 


\subsection{Displayed Rating vs Time-Averaged Aggregated Regret}
\label{sec:display-vs-aggregated-regret}


Based on the data from all the 3300 scenarios, we could not find a good  correlation between the displayed rating and time-averaged aggregated regret (no regression model gave a good fit). We then repeated analysis after discarding the data which satisfy at least one of the following (for reasons similar to the one described in Section~\ref{sec:hypo1}): 
\begin{itemize}
\item The displayed rating was less than $4$. This was to remove the transient effect from initial condition $r^s[1]=2.5$. This accounted for only about 5 \% of total data points. 
\item The time-averaged aggregated regret (cf. \eqref{eq:subject-model}) is less than $2.6$ minutes. This was because the displayed ratings varied vastly during such scenarios. This accounted for about 50 \% of total data points. 
\end{itemize}

\begin{figure}[htb!]
\begin{center}
\begin{minipage}[c]{.3\textwidth}
\begin{center}
\includegraphics[width=1.0\textwidth]{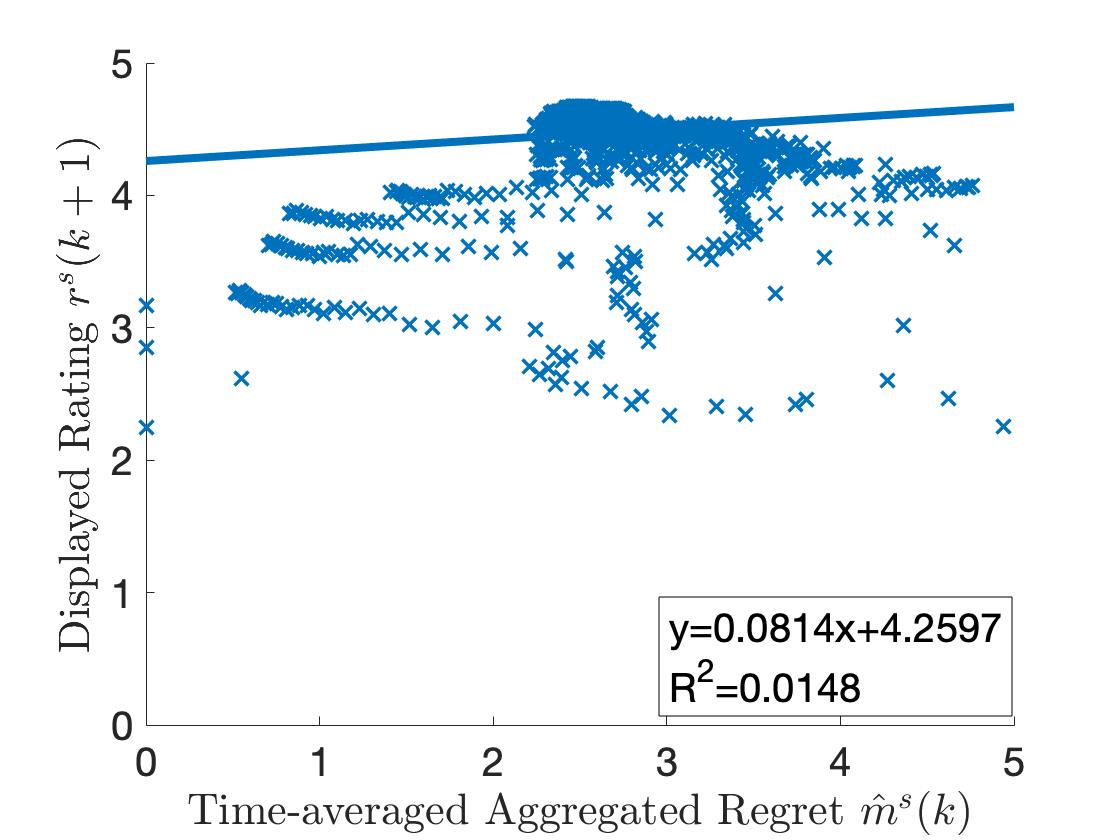} 
(a) \, (0, 2.25, 100\%)
\end{center}
\end{minipage}
\begin{minipage}[c]{.3\textwidth}
\begin{center}
\includegraphics[width=1.0\textwidth]{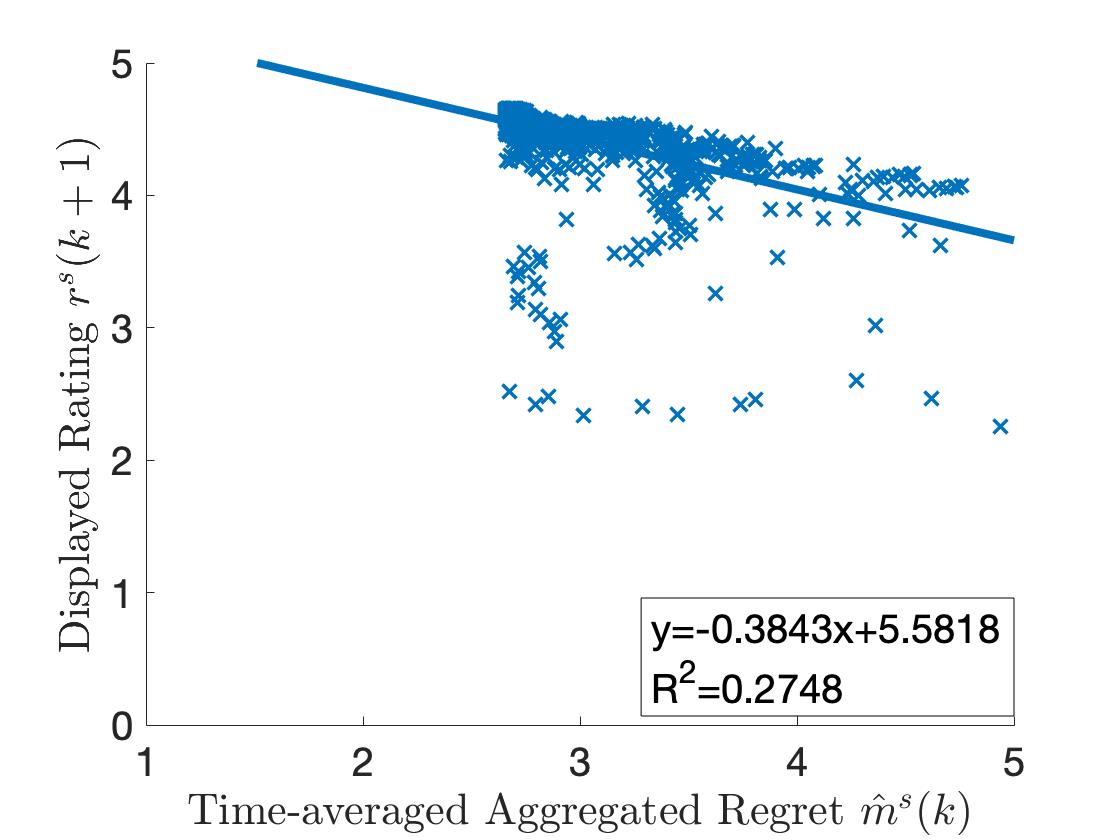} 
(b) \, (2.6, 2.25, 46\%)
\end{center}
\end{minipage}
\begin{minipage}[c]{.3\textwidth}
\begin{center}
\includegraphics[width=1.0\textwidth]{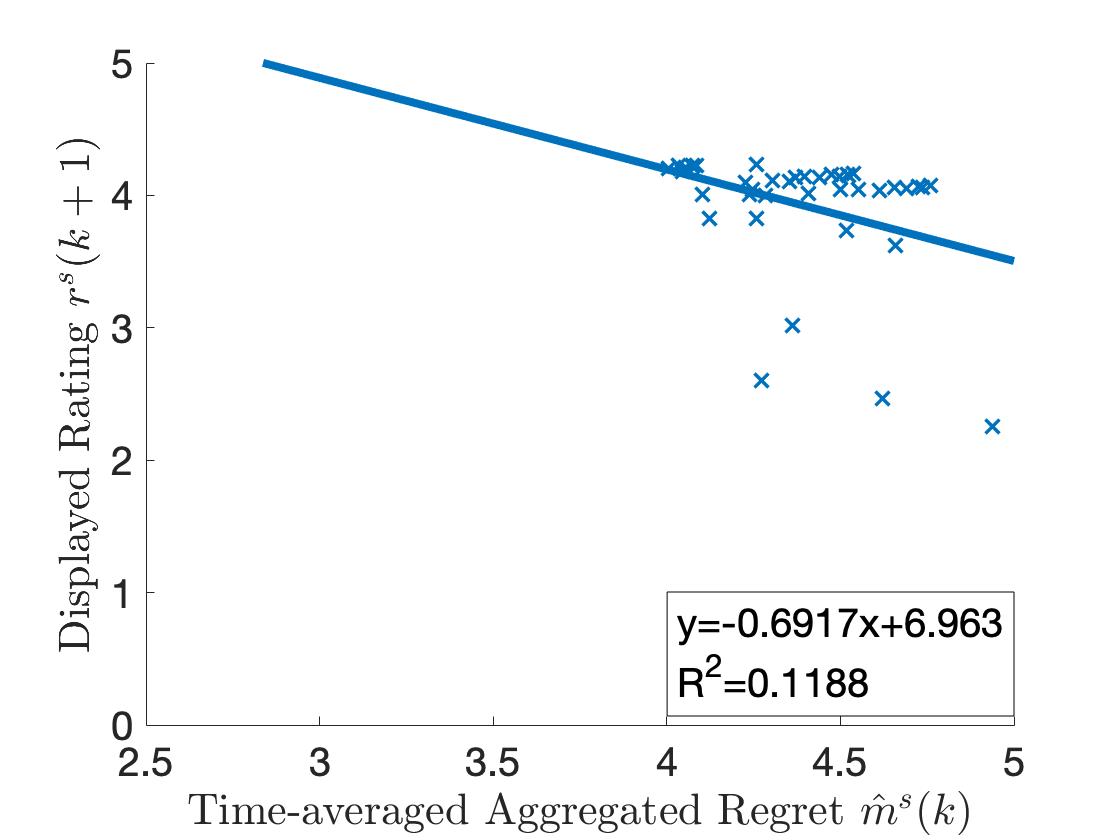} 
(c) \, (4, 2.25, 1\%)
\end{center}
\end{minipage}
\end{center}
\begin{center}
\begin{minipage}[c]{.3\textwidth}
\begin{center}
\includegraphics[width=1.0\textwidth]{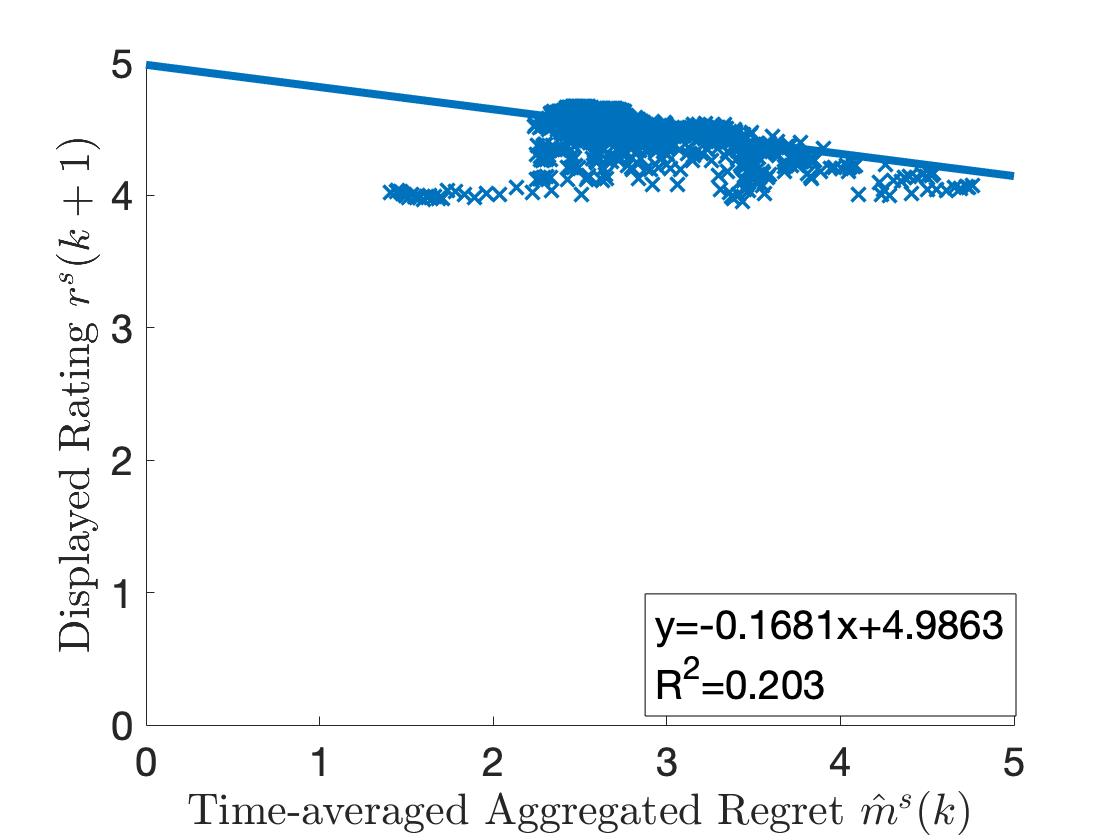}
(d) \, (0, 4, 95\%) 
\end{center}
\end{minipage}
\begin{minipage}[c]{.3\textwidth}
\begin{center}
\includegraphics[width=1.0\textwidth]{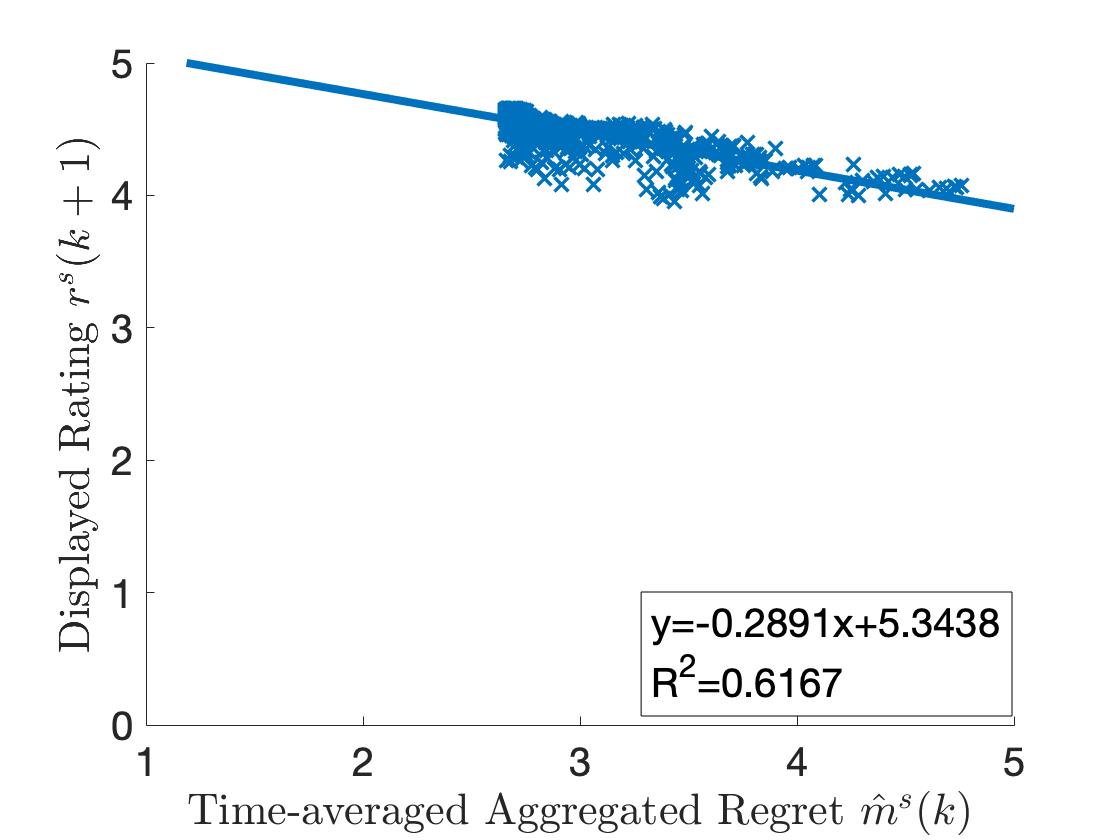} 
(e) \, (2.6, 4, 44\%)
\end{center}
\end{minipage}
\begin{minipage}[c]{.3\textwidth}
\begin{center}
\includegraphics[width=1.0\textwidth]{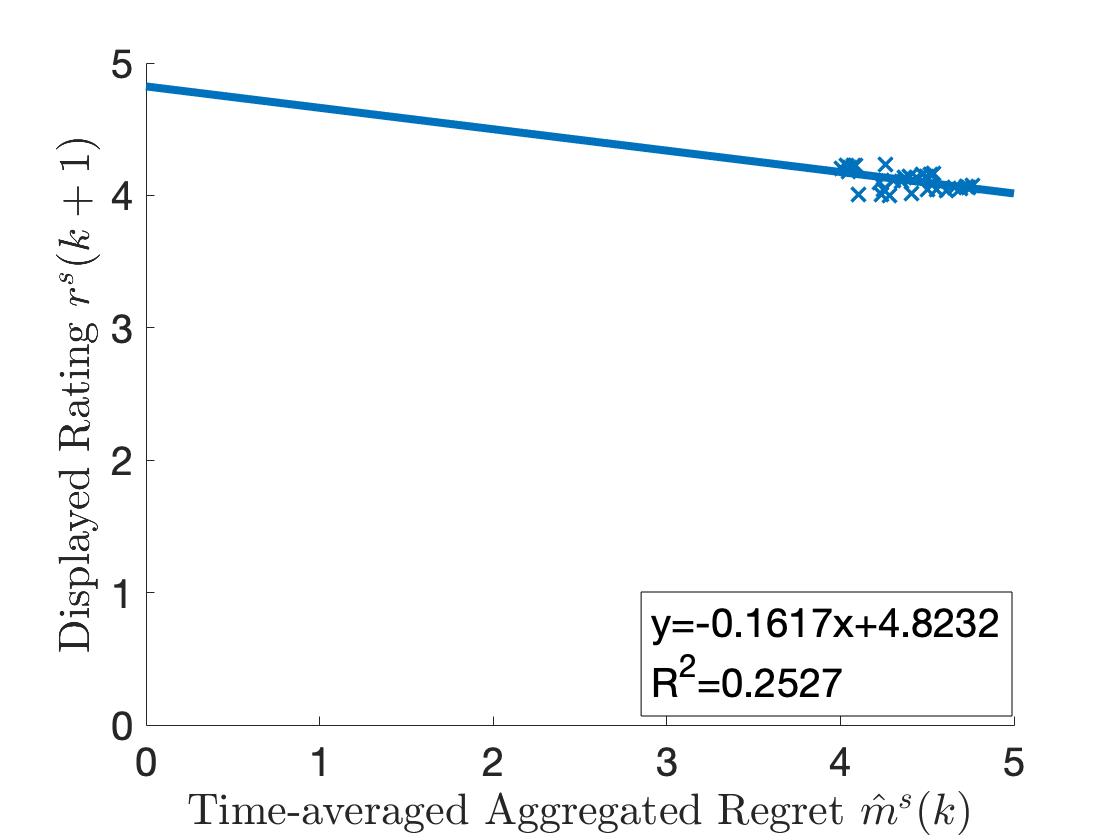} 
(f) \, (4, 4, 1\%)
\end{center}
\end{minipage}
\end{center}
\begin{center}
\begin{minipage}[c]{.3\textwidth}
\begin{center}
\includegraphics[width=1.0\textwidth]{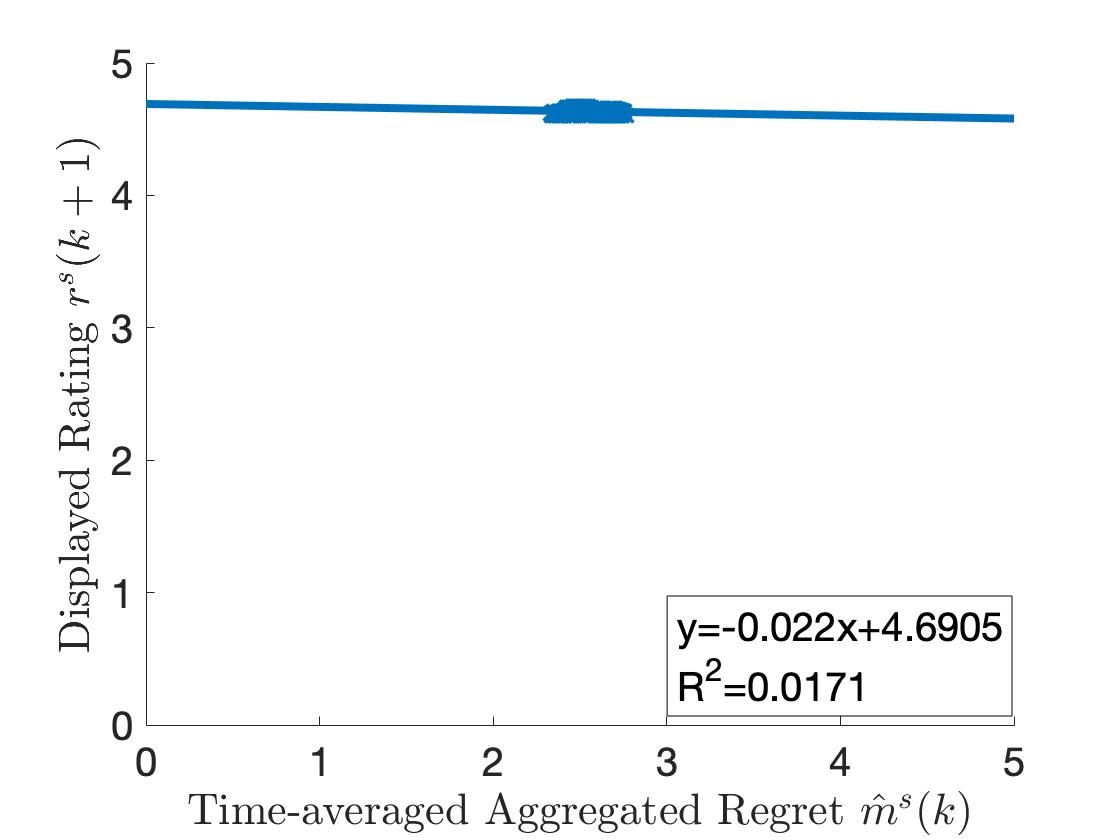} 
(g) \, (0, 4.6, 33\%)
\end{center}
\end{minipage}
\begin{minipage}[c]{.3\textwidth}
\begin{center}
\includegraphics[width=1.0\textwidth]{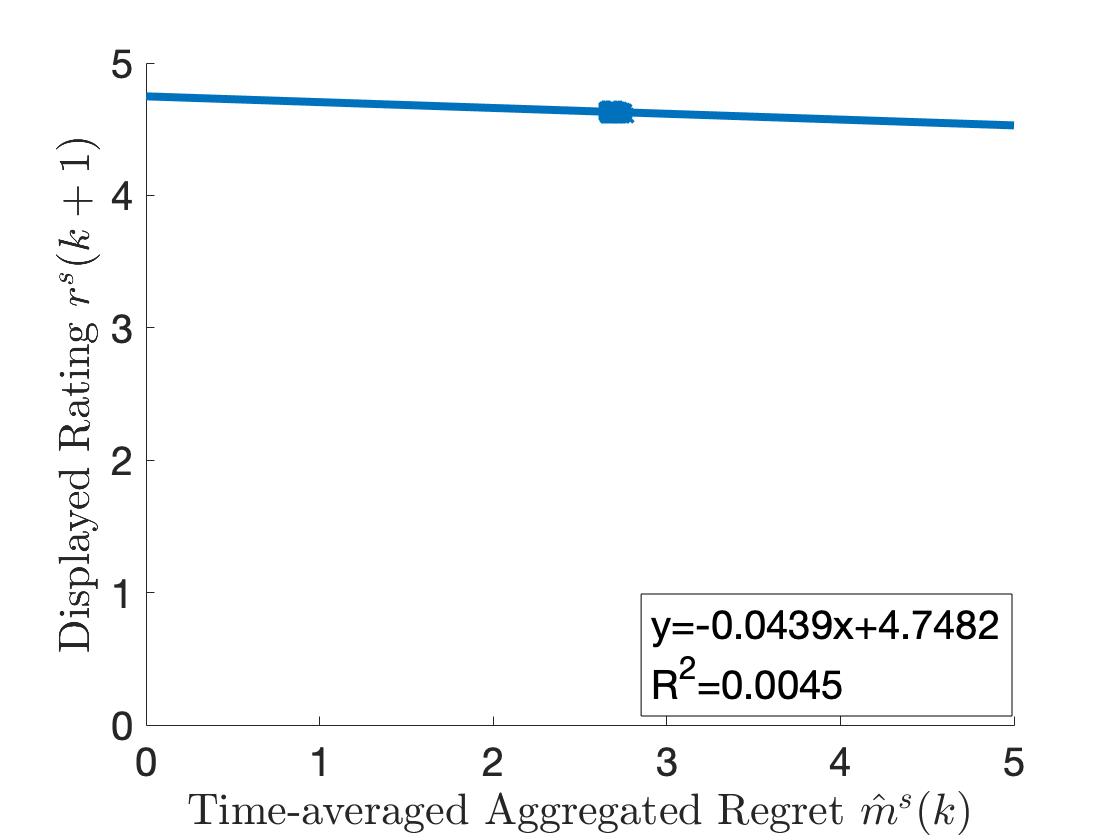} 
(h) \, (2.6, 4.6, 13\%)
\end{center}
\end{minipage}
\begin{minipage}[c]{.3\textwidth}
\begin{center}
\includegraphics[width=1.0\textwidth]{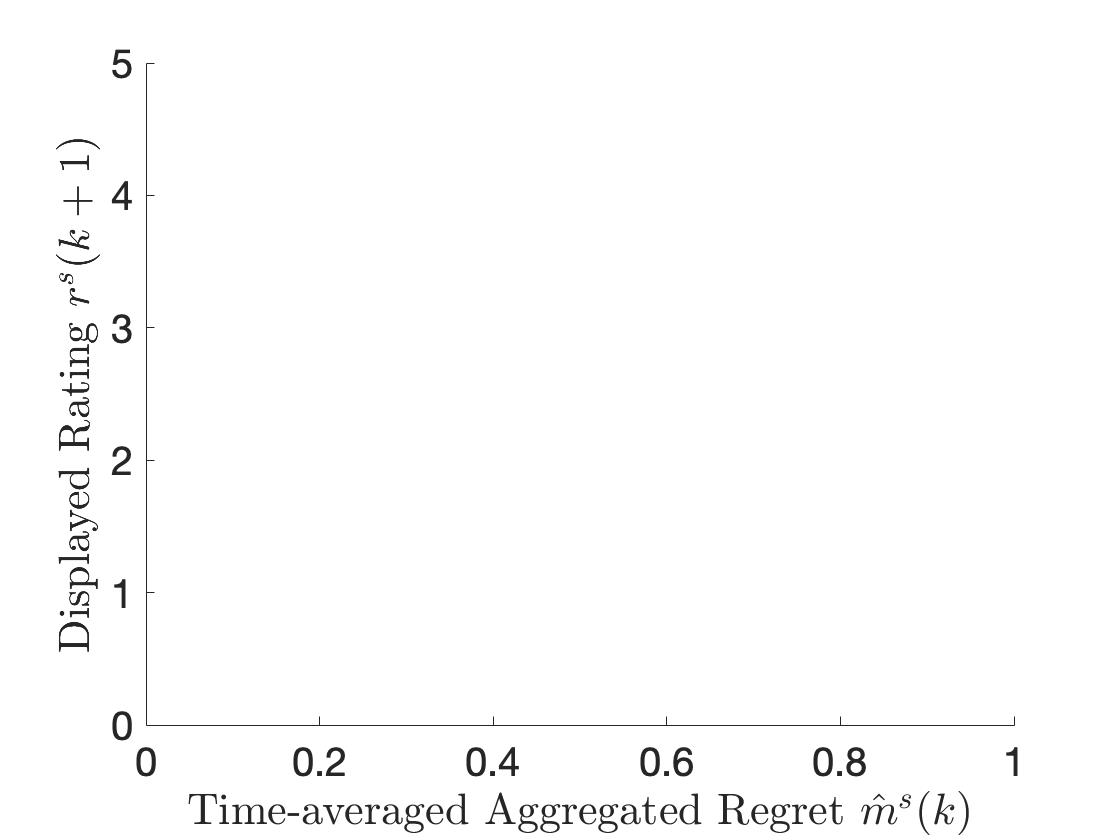} 
(i) \, (4, 4.6, 0\%)
\end{center}
\end{minipage}
\end{center}
\caption{\sf Relation between displayed rating and time-averaged aggregated regret for all participants.}
\label{fig:hypo3}
\end{figure}

%
Figure~\ref{fig:hypo3} (e) shows the outcome of linear regression between the displayed rating and the corresponding time-averaged aggregated regret values after pre-processing the data using aforementioned threshold values. The $R^2=0.6167$ value suggests a moderate negative correlation. Figure~\ref{fig:hypo3} also shows linear regression outcomes using different combinations of threshold values. For example, Figure~\ref{fig:hypo3} (f) shows the regression after discarding data with displayed rating less than 4, or when the time-averaged aggregate regret is less than 4 minutes, and that this resulted in retaining 1\% of total data points. The scenario in Figure~\ref{fig:hypo3} (e) gives the maximum $R^2$ value. 

We also investigated correlation on an individual participant basis. We found that a large number of participants submitted $\Rmax=5$ review too often. Indeed, only 9 out of the 33 participants submitted review of $5$ less than half of their individual 100 scenarios. This might be because the default review on the slider is set to $5$ before the participant adjusts it to submit his/her own review, and hence data from participants who submit a review $5$ too often might be biased. 



\begin{figure}[htb!]
\begin{center}
\begin{minipage}[c]{.475\textwidth}
\begin{center}
\includegraphics[width=1.0\textwidth]{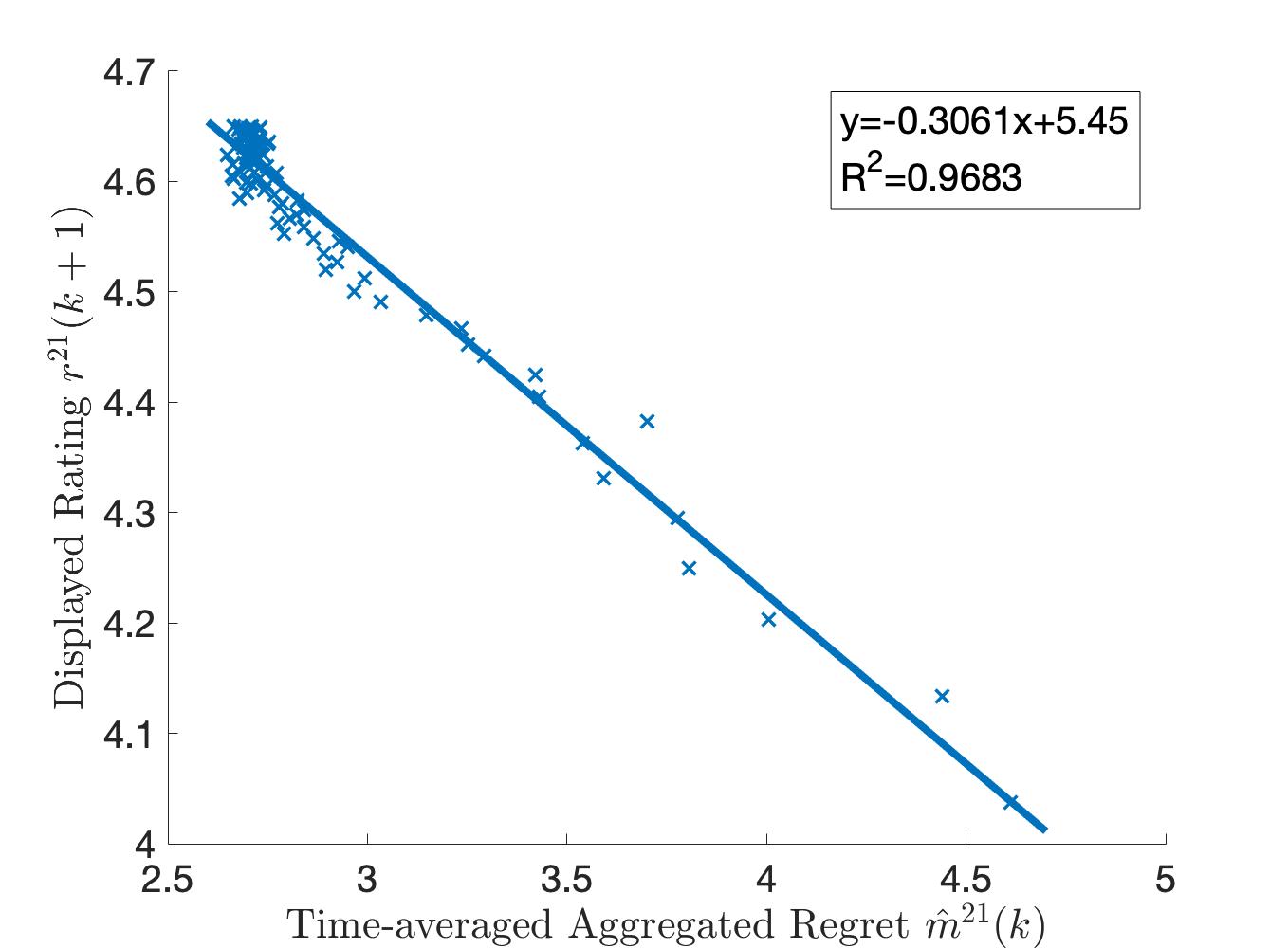} 
\end{center}
\end{minipage}
\end{center}
\caption{\sf Relation between displayed rating and time-averaged aggregated regret for a sample participant.}
\label{fig:hypo3_22}
\end{figure}

Analysis of data on an individual basis from $9$ participants who submitted review of $5$ less than half times revealed that, for $5$ of such participants, there was a strong negative correlation ($R^2>0.93$) between displayed rating and time-averaged aggregated regret. This relationship for one such participant ($\#21$) is shown in Figure~\ref{fig:hypo3_22}, where we discarded data points ($\approx 4\%$ of total data) when the review rating was less than $4$ to remove the transient effect from initial condition $r^{21}(1)=2.5$. The average $R^2$ value\footnote{The \emph{average $R^2$ value} for multiple participants refers to the average of the $R^2$ values computed for individual participants.} for all the $9$ participants who submitted review of $5$ less than half times was $0.649$. On the other hand, the average $R^2$ value for the other $24$ participants who submitted review of $5$ more than half times was $0.4664$.

We repeated our analysis by taking into account participants' response to feedback survey, specifically to questions 6 and 7 (see Appendix). $12$ participants self-reported to be using the  strategy suggested in pre-experiment instructions for making route choice decision by carefully taking into account recommended route, displayed rating and histograms. The remaining $21$ participants self-reported to be following recommendations if the displayed rating is above a threshold value. The average of this self-reported threshold value was $4.175$.

We analyzed the data separately for these two types of participants. After discarding data points corresponding to displayed rating less than $4$, analysis of the remaining data from the $12$ participants who used the strategy suggested in the instructions revealed a moderate negative correlation between displayed rating and time-averaged aggregated regret (average $R^2$ = $0.5897$). For $5$ of such participants, there was a strong negative correlation (average $R^2>0.93$). On the other hand, for the rest $21$ participants who used threshold based strategy, their average $R^2$ value was $0.4708$. 

There were $2$ participants who self-reported to use the recommended strategy, and also submitted review of $5$ less than half times. The average $R^2$ value for these $2$ participants was $0.9505$, suggesting a strong (negative) correlation between displayed rating and time-averaged aggregated regret for them.




\section{Conclusion and Future Work}
\label{sec:conclusion}
%
The immediate utility of the proposed experimental setup is to provide a setting which induces high rate of following of private route recommendation in traffic networks with uncertain travel times. The fact that the experiment design mirrors setup for a learning model with provable convergence to the Bayes correlated equilibrium induced by the obedient policy in our prior work in \cite{Zhu.Savla:DGAA21} gives further credence to this setting. In addition to providing evidence for overall convergence, our analysis also provides insight into empirical validity of key components of the learning model. Specifically, we found strong correlation between the likelihood of a particular agent following recommendation with the summary statistic of experience of other agents encapsulated by the displayed rating. The empirical route choice distributions conditional on recommendation also seem to converge. The fact that we found only moderate correlation in general between the average regret and the displayed rating, and that several participants self-reported to follow the recommendation even for medium values of displayed rating suggests further investigation is needed to conclusively establish regret as a key driver in learning. Our experimental strategy to approximate non-atomic games with atomic settings could be of independent interest for experimental studies in non-atomic games, especially with regards to correlated equilibria.  

Directions for testing robustness of the experimental findings reported in this paper include randomizing initial condition on display rating for participants as well as the default value on the slider used for collecting review from the participants, and relaxing explicit suggestions to the participants for following a specific route choice strategy. It would also be interesting to repeat the experiments under an obedient policy which induces a \emph{bad} correlated equilibrium~\cite{Duffy2010}, i.e., which is (Pareto) inferior to the Bayes Wardrop equilibrium. 
It would also be interesting to explore connection between our agent model in \eqref{eq:instant}-\eqref{eq:p flow dynamics-m-general-matrix} and the learning models in \cite{Arifovic.Ledyard:11,Arifovic.Boitnott.ea:19}.

\section*{Acknowledgment}
The authors thank Christine Stavish for help with writing code for the experiment interface and for setting up data management. The authors also thank Christine Stavish, Grace Foltz, and Alejandra Reyes for help with conducting the experiments. 

\bibliographystyle{ieeetr}
\bibliography{ksmain-exp,savla-exp}

\begin{thebibliography}{10}

\bibitem{Wang.Li.ea:14}
H.~Wang, G.~Li, H.~Hu, S.~Chen, B.~Shen, H.~Wu, W.-S. Li, and K.-L. Tan, ``R3:
  a real-time route recommendation system,'' {\em Proceedings of the VLDB
  Endowment}, vol.~7, no.~13, pp.~1549--1552, 2014.

\bibitem{Herzog.Massoud.ea:17}
D.~Herzog, H.~Massoud, and W.~W{\"o}rndl, ``Routeme: A mobile recommender
  system for personalized, multi-modal route planning,'' in {\em Proceedings of
  the 25th Conference on User Modeling, Adaptation and Personalization},
  pp.~67--75, 2017.

\bibitem{Bergemann.Morris:19}
D.~Bergemann and S.~Morris, ``Information design: A unified perspective,'' {\em
  Journal of Economic Literature}, vol.~57, no.~1, pp.~44--95, 2019.

\bibitem{Zhu.Savla:DGAA21}
Y.~Zhu and K.~Savla, ``Convergence analysis for repeated non-atomic games with
  partial signaling.'' Working draft available at
  \texttt{https://viterbi-web.usc.edu/$\sim$ksavla/papers/convergence-partial-signaling.pdf},
  2021.

\bibitem{Fischer.Racke.ea:10}
S.~Fischer, H.~R{\"a}cke, and B.~V{\"o}cking, ``Fast convergence to wardrop
  equilibria by adaptive sampling methods,'' {\em SIAM Journal on Computing},
  vol.~39, no.~8, pp.~3700--3735, 2010.

\bibitem{Blum.Even-Dar.ea:10}
A.~Blum, E.~Even-Dar, and K.~Ligett, ``Routing without regret: On convergence
  to nash equilibria of regret-minimizing algorithms in routing games,'' {\em
  Theory of Computing}, vol.~6, no.~1, pp.~179--199, 2010.

\bibitem{Krichene.Drighes.ea:15}
W.~Krichene, B.~Drigh{\`e}s, and A.~M. Bayen, ``Online learning of nash
  equilibria in congestion games,'' {\em SIAM Journal on Control and
  Optimization}, vol.~53, no.~2, pp.~1056--1081, 2015.

\bibitem{Iida.Akiyama.ea:92}
Y.~Iida, T.~Akiyama, and T.~Uchida, ``Experimental analysis of dynamic route
  choice behavior,'' {\em Transportation Research Part B: Methodological},
  vol.~26, no.~1, pp.~17--32, 1992.

\bibitem{Selten.Chmura.ea:07}
R.~Selten, T.~Chmura, T.~Pitz, S.~Kube, and M.~Schreckenberg, ``Commuters route
  choice behaviour,'' {\em Games and Economic Behavior}, vol.~58, no.~2,
  pp.~394--406, 2007.

\bibitem{Cason.Sharma:07}
T.~N. Cason and T.~Sharma, ``Recommended play and correlated equilibria: an
  experimental study,'' {\em Economic Theory}, vol.~33, no.~1, pp.~11--27,
  2007.

\bibitem{Duffy2010}
J.~Duffy and N.~Feltovich, ``Correlated equilibria, good and bad: An
  experimental study,'' {\em International Economic Review}, vol.~51, no.~3,
  pp.~701--721, 2010.

\bibitem{Bone.Drouvelis.ea:12}
J.~Bone, M.~Drouvelis, and I.~Ray, {\em Following Recommendation to Avoid
  Coordination-Failure in 2 x 2 Games}.
\newblock Department of Economics, University of Birmingham, 2012.

\bibitem{Duffy2017}
E.~K.~L. John~Duffy and W.~Lim, ``Coordination via correlation: An experimental
  study,'' {\em Economic Theory}, vol.~64, pp.~265--304, 2017.

\bibitem{Arifovic.Boitnott.ea:19}
J.~Arifovic, J.~F. Boitnott, and J.~Duffy, ``Learning correlated equilibria: An
  evolutionary approach,'' {\em Journal of Economic Behavior \& Organization},
  vol.~157, pp.~171--190, 2019.

\bibitem{Anbarci2018}
N.~F. Nejat~Anbarc$\text{\i}$ and M.~Y. G$\ddot{\text{u}}$rdal, ``Payoff
  inequity reduces the effectiveness of correlated-equilibrium
  recommendations,'' {\em European Economic Review}, vol.~108, pp.~172--190,
  2018.

\bibitem{Zhu.Savla:TCNS22}
Y.~Zhu and K.~Savla, ``Information design in non-atomic routing games with
  partial participation: Computation \& properties,'' {\em IEEE Transactions on
  Control of Network Systems}, 2022.

\bibitem{Hart.Mas-Colell:00}
S.~Hart and A.~Mas-Colell, ``A simple adaptive procedure leading to correlated
  equilibrium,'' {\em Econometrica}, vol.~68, no.~5, pp.~1127--1150, 2000.

\bibitem{RoutingExpSlides}
``Participant instruction slides for the experiment \textit{Long-term Route
  Choice Decisions under Personalized Recommendations}.''
\newblock Available at
  \url{https://viterbi-web.usc.edu/~ksavla/misc/OrientationSlides.pptx}.

\bibitem{Arifovic.Ledyard:11}
J.~Arifovic and J.~Ledyard, ``A behavioral model for mechanism design:
  Individual evolutionary learning,'' {\em Journal of Economic Behavior \&
  Organization}, vol.~78, no.~3, pp.~374--395, 2011.

\end{thebibliography}

\appendix

\subsection{Feedback Survey}
\begin{enumerate}
\item On the scale of 1 to 5 (with 5 being the highest), how did you feel you understood what was happening in each scenario, what you were doing and what you needed to do next?
\item On the scale of 1 to 5 (with 5 being the highest), how often did you check the average star rating before making route choice decisions?
\item When you indeed checked the average star rating, on the scale of 1 to 5 (with 5 being the highest), how much did the average star rating affect your route choice decisions?
\item On the scale of 1 to 5 (with 5 being the highest), how often did you check the histograms before making route choice decisions?
\item When you indeed checked the histograms, on the scale of 1 to 5 (with 5 being the highest), how much did the histograms affect your route choice decisions?
\item How did you make your route choice decisions?
\begin{enumerate}
\item Study the average star rating, histograms as well as the recommendations to come up with a decision.
\item Follow the recommendations as long as the average star rating is above certain value.
\item Always follow the recommendations.
\item Make random choices.
\end{enumerate}
\item If you choose (b) in question 6, what was the threshold value? 
\item What did you like (if any) or dislike (if any) about the experiment interface?
\item Any other comments you might have
\end{enumerate}


\end{document}